%% file: main.tex
\documentclass[%
  twocolumn,
groupedaddress,
showpacs,preprintnumbers,
nofootinbib,
 amsmath,amssymb,
 aps,
 prd,
]{revtex4-1}

\usepackage{braket}            %for braket notation
\usepackage{graphicx}          %Include figure files
\usepackage[utf8]{inputenc}    %For Bali paper references
\usepackage{hyperref}          %Hyperlinks for references, cites, etc
\usepackage{dcolumn}           % Align table columns on decimal point
\usepackage{color}             %for \textcolor
\usepackage{adjustbox}         %For making the chiEFT coeffs table fit better
\usepackage[caption=false]{subfig}        %Subfigures
\usepackage{csquotes}        %enquote
\usepackage{physics}
\usepackage{booktabs}
\usepackage{amsmath}
\usepackage{animate}

\makeatletter
\let\oldtheequation\theequation
\renewcommand\tagform@[1]{\maketag@@@{\ignorespaces#1\unskip\@@italiccorr}}
\renewcommand\theequation{(\oldtheequation)}
\makeatother

\makeatletter
\def\maketitle{
	\@author@finish
	\title@column\titleblock@produce
	\suppressfloats[t]}
\makeatother

\newcommand{\mpipacs}{m_{\pi}^{PACS}}

\newcommand{\sigmap}{$\Sigma^+$}
\newcommand{\sigmam}{$\Sigma^-$}
\newcommand{\sigmaz}{$\Sigma^0$}
\newcommand{\cascadez}{$\Xi^0$}
\newcommand{\cascadem}{$\Xi^-$}

\begin{document}

\preprint{ADP-24-22/T1261}

\title{Magnetic polarisability of octet baryons near the physical quark-mass point}
\author{Thomas Kabelitz}
\author{Waseem Kamleh}
\author{Derek Leinweber}

\affiliation{Special Research Centre for the Subatomic Structure of Matter (CSSM),\\
    Department of Physics, University of Adelaide, Adelaide, South Australia 5005, Australia
}

\date{\today}

\begin{abstract}
    The magnetic polarisabilities of octet baryons are calculated close to the physical quark-mass
    point using the background field method in lattice QCD. This first calculation draws on the
    identification and elimination of exceptional configurations that have hindered previous attempts.
    The origin of the exceptional configuration problem lies in the use of a Wilson-type fermion action
    on electro-quenched gauge field configurations, where the dynamical-fermion gauge-field generation
    algorithm the electric charges of the quarks.
    Changes in the fermion determinant that would suppress some gauge fields in the background magnetic
    field are neglected, leaving improbable gauge fields that generate large additive mass
    renormalisations which manifest as significant outliers in correlation-function distributions.
    An algorithm for the systematic identification and removal of these exceptional configurations is
    described. We find the light up and down quarks to be problematic, particularly the up quark
    with its larger electric charge.
    The heavier mass of the strange quark protects the hyperon correlation functions to some extent.
    However, these also benefit from the removal of exceptional configurations.
    In many cases, the magnetic polarisability is calculated with good precision.
    We find our results to be in accord with the behaviour anticipated by chiral perturbation theory.
\end{abstract}

%13.40-f = Electromagnetic processes and properties
%12.38.Gc = Lattice QCD calculations
%12.39.Fe = Chiral Lagrangians
%\keywords{}

\maketitle

\section{Introduction}

The uniform background magnetic field formalism has been effective in enabling precision lattice
QCD calculations of hadron magnetic moments and magnetic polarisabilities. By measuring the change
in the energy of the hadron as a function of the background field strength, one can extract these
quantities with uncertainties that compete favourably with experimental measurements.

However, calculations of the magnetic
polarisability~\cite{hall2014chiral,bignell2018neutron,bignell2020nucleon,bignell:2019:cloverpion,
    kabelitz:octet} have previously been unable to probe the light near-physical quark-mass regime
due to large and uncontrolled statistical uncertainties that appear in lattice two-point
correlation functions at finite magnetic field. Statistical distributions do not respond to
increases in the number of samples in the expected manner, indicating an important systematic
error.

To make contact with experiment, chiral effective field theory has been used to extrapolate
lattice calculations from larger unphysical quark masses to the physical point. At light quark
masses, significant chiral nonanalytic behaviour is predicted.
Whether this behaviour is observed in lattice QCD calculations remains to be determined.

In this work we highlight the presence of an exceptional configuration problem encountered at
near-physical quark masses in nontrivial uniform background magnetic fields.
The origin of the exceptional configuration problem lies in the use of a Wilson-type fermion action
on electro-quenched gauge field configurations. Here, the dynamical-fermion gauge-field generation
is performed at zero magnetic field, effectively neutralising the electric charges of the quarks.
In other words, the sea quarks are blind to the background magnetic field. Changes in the
fermion determinant that would suppress some gauge fields are neglected. These improbable gauge
fields can generate large additive mass renormalisations which manifest as significant outliers in
correlation-function distributions.

We show how these exceptional configurations render the extraction of signal for nucleons
impossible. While the heavier strange quarks present in the hyperons allows the extraction of some
signal, the exceptional configurations in the light sector do impact the quality of that signal.

Our main focus is to develop and test algorithms to identify these exceptional configurations and
remove them from lattice QCD correlation functions.  This enables a high statistics calculation of
the magnetic polarisability of ground-state octet baryons at near physical quark masses on the
lightest PACS-CS ensemble~\cite{PACS-CS2008ensembles}. We utilise the background field method
and calculate the magnetic polarisability through a combination of spin-field aligned and
anti-aligned two-point correlation functions. As the $\Lambda$ and $\Sigma^0$ baryons mix in the
background field, we consider the outer octet baryons composed of a doubly represented quark
flavour and a singly represented quark flavour.

In \autoref{sec:background} we briefly discuss the setup of the calculation and the simulations
details are reviewed in \autoref{sec:simulationdetails}.
The exceptional configuration problem is addressed in \autoref{sec:exceptionalconfigurations},
where algorithms for the systematic identification and removal of exceptional configurations are
created and implemented.  Animations showing the evolution of lattice correlation functions as
exceptional configurations are eliminated are presented in the Supplementary Information.
Following the removal of exceptional configurations, octet-baryon magnetic polarisabilities are
calculated at the lightest PACS-CS quark mass with $m_\pi = 156$ MeV and presented in
\autoref{sec:latticeresults}.
Drawing on our previous results at larger quark masses in Ref.~\cite{kabelitz:octet}, chiral
extrapolations are performed in \autoref{sec:chiralextrapolation} and compared with our new results
to evaluate the chiral behaviour observed in lattice QCD
Finally, a summary of our findings is presented in \autoref{sec:conclusion}.

\section{Calculation Setup}\label{sec:background}

Our approach to the extraction of the baryon magnetic polarisability follows our approach in
Ref.~\cite{kabelitz:octet}.
In this section, we briefly summarise the approach, and full details are available in
Ref.~\cite{kabelitz:octet}.

In the uniform background field method, a minimal coupling to the discretised covariant derivative
results in the application of a phase to the QCD gauge field. Through the requirement for
periodicity at the two boundaries orthogonal to the magnetic field direction, a
quantisation condition is obtained for the magnetic field strength $B$, given by
\begin{equation}\label{eqn:backgroundfield:quantisationcondition}
    e\,B=\frac{2\,\pi}{N_x\,N_y\,a^2}\frac{1}{q_d}\,k_d,
\end{equation}
where $k_d$ is an integer specifying the field strength in multiples of the minimum uniform field
strength quantum. The condition is written in terms of the down quark charge $q_d\,e$, the number
of spatial sites $N_x,\,N_y$ and the lattice spacing $a$.

The magnetic field is defined such that the field corresponding to integer $k_d=1$ is oriented in
the negative $\hat{z}$-direction. This work utilises quark propagators and baryon correlation
functions calculated at \mbox{$k_d=0,\,\pm 1,\, \pm 2$}. We will refer to the strength of the magnetic
field in terms of $k_d$ throughout this work.

This uniform background magnetic field imparts a change in the energy of the baryon as a
function of magnetic field strength \cite{Martinelli:1982:expansion,primer2014magnetic}
\begin{equation}\label{eqn:backgroundfield:energyexpansion}
    E(B) = m + \vec{\mu}\cdot\vec{B} + \frac{|q_B\,e\,B|}{2\,m}\left(n+1\right) -
    \frac{1}{2}\, 4 \pi\,\beta\,B^2 + \order{B^3} \, .
\end{equation}
Here the mass of the baryon, $m$, is complemented by contributions from the magnetic moment
$\vec{\mu}$, the Landau term proportional to $|q_B\, e\, B|$ where $q_B$ is the charge of the
baryon, and the magnetic polarisability $\beta$. We use a U(1) Landau-mode
projection~\cite{bignell2020nucleon} to select $n=0$.
% See Ref.~\cite{kabelitz:octet} for details.

The sign of the magnetic moment term is dependent on the alignment of the baryon's spin and the magnetic field. As such, it may be eliminated through summation of a \emph{spin-field aligned} correlation function $G_{\uparrow\uparrow}(B)$ and a \emph{spin-field anti-aligned} correlation function $G_{\uparrow\downarrow}(B)$.

The effective energy of the correlation function ratio
\begin{equation} \label{eqn:polarisability:ratio}
    R(B,t) = \frac{G_{\uparrow\uparrow}(B,t)\,G_{\uparrow\downarrow}(B,t)}{G(0,t)^2} \, ,
\end{equation}
subtracts the magnetic moment term and the contribution from the baryon mass leaving the magnetic polarisability energy shift $\delta E_{\beta}(B,t)$
\begin{align}
    \delta E_{\beta}(B,t) & = \frac{1}{2}\frac{1}{\delta t}\lim_{t\rightarrow
        \infty}\log\Bigl(\frac{R(B,t)}{R(B,t+\delta t)}\Bigr) \, ,
    \label{eqn:polarisability:energy:ratio}
    \\
    & = \frac{\abs{q\, e\, B}}{2m} - \frac{4\pi}{2}\beta|B|^2 +\order*{B^3}\, .
    \label{eqn:polarisability:energyshift}
\end{align}

As the construction of the ratio $R(B,t)$ aggregates the positive and negative magnetic field strengths, we fit $\delta E_{\beta}(B,t)$ as a function of field strength to extract the magnetic polarisability.

In practice, it is simpler to fit in terms of the field strength quanta $k_d$. Using the quantisation condition Eq.~\ref{eqn:backgroundfield:quantisationcondition}, we can write the above as
\begin{equation}
    \delta E_{\beta}(k_d) = L(k_d, m) + C \, \beta \, k_d^2 \, , \label{eqn:fitting:fieldstrengthfit}
\end{equation}
where the Landau term is given by
\begin{equation}
    L(k_d, m) = \frac{2\pi}{N_x\, N_y \, a^2}\frac{1}{2 \, m}\abs{\frac{q_B}{q_d} \, k_d},
\end{equation}
and the coefficient to the $k_d^2$ term is
\begin{equation}
    C = - \frac{1}{2 \, \alpha}\left[\frac{2\pi}{N_x \, N_y \, a^2}\right]^2\frac{1}{q_d^2} \, . \\
\end{equation}

Here $\alpha=\frac{e^2}{4\pi}$ is the fine structure constant. It is important to note that $q_d$ is the down quark charge while $q_B$ is the hadronic charge of the baryon.

To obtain $\delta E_{\beta}(k_d)$ and $m$, we utilise a weighted average of fit windows to provide
a systematic approach to fitting. We consider fit windows $[t_s,t_f]$ with minimum length three
corresponding to $t_f\geq t_s +2$ in range $[t_{\rm min},t_{\rm max}]$. We set $t_f=t_{\rm max}$ to
ensure equitable sampling of all time slices. $t_{\rm max}$ is chosen as the last time slice before
signal is lost to noise. We choose $t_{\rm min} = 20$ as the level of available signal we have
access to at the lightest PACS-CS quark mass makes it impossible to consistently satisfy the
``$\log G$'' requirements we imposed at heavier quark masses in Ref.~\cite{kabelitz:octet}. The
choice of $t_{\rm min} = 20$ is chosen to give maximum opportunity for the excited state
contamination to ease while ensuring that we have windows of the requisite length available for
sampling.
The utilisation of a weighted average of eligible windows enforces the sampling of later windows
and this will further aid in suppressing the excited state contamination.

In Ref.~\cite{kabelitz:octet}, we utilised a criterion which weighs the probability of
the fit's $\chi^2$ compared to its uncertainty. This weighting is appropriate when there is already
confidence that all candidate windows have reached single-state isolation. As we are unable to
consistently fit in such a region on this near-physical mass ensemble, we utilise a different
weighting criterion. Here we use a form of the Akaike information criterion (AIC)~\cite{Akaike:1974vps} to provide a
weight ~\cite{Borsanyi:2020mff:AICweight}
\begin{equation}
    w = \exp\left(-\frac{1}{2}\left[\chi^2 + n_{\rm data} - 2\,n_{\rm d.o.f.}\right]\right),
\end{equation}
for each fit window.  This measure depends purely on the quality of the fit. This form of the AIC
inherently biases later, shorter windows, further suppressing any excited state contamination.

The weights are normalised, $\tilde{w}_i$, such that $\sum_i \tilde{w}_i = 1$. The average effective
energy is then
\begin{equation}
    E = \sum_i^n \tilde{w}_i \, E_i \, ,
\end{equation}
and the statistical error
\begin{equation}
    (\delta E)^2 = \sum_i^n \tilde{w}_i(\delta E_i)^2\, .
\end{equation}

A weighted average is also used to obtain the mass in the Landau term where required in fitting
Eq.~\ref{eqn:fitting:fieldstrengthfit}.
As in Ref.~\cite{kabelitz:octet}, a jackknife error estimate is performed to determine the
uncertainty in the magnetic polarisability.

\section{Simulation Details}\label{sec:simulationdetails}

The simulation details in this work are almost identical to those of our previous work
\cite{kabelitz:octet}, exempting \autoref{sec:simulationdetails:statistics} on statistics.  Here, a
much greater statistical sample is required to obtain signal in correlation functions this close to
the physical point.

\subsection{Gauge ensembles}

\begin{table}[tb]
    \centering
    \caption{Details of the PACS-CS ensembles used in this work. The lattice spacing of each
        ensemble is set using the Sommer scale with $r_0=0.4921(64)(+74)(-2)\,$fm. In all cases
        $\kappa_s^{sea}=0.13640$ and $\kappa_s^{val.}=0.13665$~\cite{menaduethesis}. $N_{\rm
                    con}$ describes the number of
        configurations.} \label{tab:simulationdetails:pacsensembles}
    \begin{ruledtabular}
        \begin{tabular}{cccc}
            $\mpipacs$(MeV) & $\kappa_{u\, d}$ & $a\,$(fm)  & $N_{\rm con}$ \\
            \noalign{\smallskip}\hline\noalign{\smallskip}
            701             & 0.13700          & 0.1022(15) & 399           \\
            570             & 0.13727          & 0.1009(15) & 397           \\
            411             & 0.13754          & 0.0961(13) & 449           \\
            296             & 0.13770          & 0.0951(13) & 399           \\
            156             & 0.13781          & 0.0933(13) & 197           \\
        \end{tabular}
    \end{ruledtabular}
\end{table}

In this work we present new results on the lightest PACS-CS ensemble while drawing on our results on the four heavier PACS-CS ensembles from Ref.~\cite{kabelitz:octet}.
The ensembles are $2+1$-flavour dynamical gauge
configurations provided by the PACS-CS collaboration~\cite{PACS-CS2008ensembles} through the
International Lattice Data Grid (ILDG)~\cite{ILDG}. The configurations have a range of degenerate
up and down quark masses while the strange quark mass is fixed. The strange quark mass of the
ensembles which corresponds to $\kappa_s=0.13640$ does not extrapolate to the physical kaon
mass~\cite{menadue2012lambda}. Use of $\kappa_s=0.13665$ for the valence strange quark mass
produces the correct value for the kaon mass extrapolated to the physical
point~\cite{menaduethesis}. We note that the mass of the strange quarks in the sea remain at the
heavier mass.

Each of the ensembles are a $32^3\times 64$ lattice. The gauge action is the Iwasaki gauge action
and the clover fermion action with $C_{\rm SW}=1.715$ is the background field corrected clover
fermion action~\cite{bignell:2019:cloverpion}, tuned to remove the non-physical magnetic-field
induced additive mass renormalisation. Details of the ensembles are summarised in
\autoref{tab:simulationdetails:pacsensembles}.  At the lightest quark mass holding the focus of
this investigation, the spatial cubic lattice has a length $L = 3.0$ fm.

As only the valence quarks interact with the background magnetic field the ensembles are
electro-quenched. While it is possible to include the background field in the process of generating
each gauge field configuration,
this would require a separate Monte-Carlo simulation at each field strength. Such separate
simulations would remove the correlated QCD fluctuations which are otherwise efficiently removed
through the ratio in Eq.~\ref{eqn:polarisability:ratio} and vital to resolving a signal.  Without
this correlation, a very significant increase in statistics would be required.  As including the
background field in the gauge-field generation process may address the exceptional configuration
problem studied here, it will be important to consider this in a next generation calculation.

\subsection{Baryon interpolating fields}

The commonly used proton interpolating field in lattice QCD is given by~\cite{Leinweber:1990dv}
\begin{equation}
    \chi_p(x) = \epsilon^{abc} \, \left[ {u^a}^T \, C \, \gamma_5 \, d^{\,b}(x)\right]\,u^c(x) \, ,
\end{equation}
where $C$ is the charge-conjugation matrix.  The interpolating fields of the other outer
octet baryons may be easily obtained through appropriate substitution of doubly and singly
represented quark flavours. This is the form of the interpolating field used throughout
this work.

Such interpolating fields, utilised purely with traditional gauge-covariant Gaussian
smearing are ineffective at isolating the baryon ground state in a uniform background
field
\cite{primer2014magnetic,Deshmukh:2018:octet,bignell2020nucleon,Bruckmann:2017pft,Chang:2015:polarisability}. The
uniform background field breaks the spatial symmetry and Landau-mode physics presents
at both the quark and hadronic levels, and this must be accommodated.

\subsection{Quark operators}

As detailed in Ref.~\cite{kabelitz:octet}, we utilise asymmetric source and sink operators to
maximise overlap of the lowest energy eigenstates of our baryons in magnetic fields. The operators
we utilise also act to accommodate for the spatial symmetry and Landau-mode issues mentioned above.

Source and sink construction utilises links which are smeared using stout link
smearing~\cite{morningstar:2003linksmearing}. 10 sweeps with isotropic smearing parameter
$\alpha_{\rm stout}=0.1$ are applied to spatially-oriented links.

We construct the quark source using three-dimensional, gauge-invariant Gaussian
smearing~\cite{gusken:1990:smearing}. $\alpha=0.7$ is used with 250 sweeps.

The source construction is designed to provide a representation of the QCD interactions with the
intent of isolating the QCD ground state. To encapsulate the quark-level physics of the
electromagnetic and QCD interactions, we utilise the eigenmode projection techniques demonstrated in
Ref.~\cite{bignell2020nucleon}, where a comprehensive explanation of the mechanisms may be found.
A succinct summary of the approach is provided in Ref.~\cite{kabelitz:octet}.

The eigenmodes of the two-dimensional lattice Laplacian calculated on the SU(3)$\times$U(1) gauge
links are used to produce a projection operator applied to the propagator at the sink. It is shown
in Ref.~\cite{bignell2020nucleon}, that including too few modes results in a noisy hadron
correlation function in much the same manner as traditional sink smearing. As such, the number of
modes is chosen to be large enough to minimise the noise of the correlation function, but small enough
to retain the focus on the aforementioned low-energy physics. The work of
Ref.~\cite{bignell2020nucleon} found that $n=96$ modes provides balance to these two effects and we
use this here.

\subsection{Boundary conditions}

For the calculation of the quark propagators, periodic boundary conditions are used in the
spatial dimensions. To avoid signal contamination from the backward propagating states, we
use fixed boundary conditions in the temporal direction. The source is then placed at
$t=16$, one quarter of the total time-dimension length such that one is always away from
the fixed boundary by using the middle part of the lattice time dimension.

\subsection{Hadronic projection}\label{sec:simulationdetails:landauprojection}

The inclusion of the background magnetic field induces a change to the wave function of a charged
baryon~\cite{roberts:2010:protonwf}. The quark level electromagnetic physics is highlighted by the
eigenmode projection at the sink. However, we must also ensure that our operator has the
appropriate electromagnetic characteristics on the hadronic level. By projecting final-state
charged baryons to the Landau state corresponding to the lowest-lying Landau level, we ensure $n=0$
in Eq.~\ref{eqn:backgroundfield:energyexpansion}.

The lowest-lying Landau mode has degeneracy \mbox{$n=\abs{k_d\frac{q_B}{q_d}}$}. As discussed in
Ref.~\cite{kabelitz:octet}, we use a linear combination of degenerate eigenmodes providing an
optimal overlap with the quark source. This hadronic eigenmode-projected correlation function
offers superior isolation of the ground state as shown in Ref.~\cite{tiburzi:2012:projection} and
is crucial for the fitting of constant plateaus in the energy shift of
Eq.~\ref{eqn:polarisability:energyshift} herein.

\subsection{Magnetic field}

Baryon correlation functions are calculated for five magnetic-fields corresponding to
$k_d=-2,\, -1,\, 0,\, 1,\, 2$. In doing so, quark propagators and eigenmodes are
calculated at $k_d=0,\, \pm 1,\, \pm 2,$ and $\pm 4$ in accounting for the up quark. The
non-zero field strengths correspond to magnetic fields in the $z$-direction of $e\, B=\pm
    0.087$, $\pm 0.174$, and $\pm0.348\,$GeV$^2$.

\subsection{Statistics}\label{sec:simulationdetails:statistics}

As periodic boundary conditions are used in all four dimensions for the gauge field
generation, one can exploit the associated translational invariance of the gauge fields. A
quark source can be placed at any position on the lattice and then circularly
cycled to the standard source position of $(x,y,z,t)=(1,1,1,16)$. This enables additional
sampling of the full gauge field.

Further, the two-dimensional nature of the lattice Laplacian operator allows the
eigenmodes for the sink projection to be re-used when the gauge field is cycled solely in the
time direction.

Hence, we increase our statistics on the PACS-CS ensembles by considering a random
spatial source location at $t=16$ and then circularly cycling in spatial dimensions to $(1,1,1,16)$.
This spatial source is considered eight times by circularly cycling the gauge field in the temporal
direction by an eighth of the lattice time extent (eight slices in our case).

On the four heavier PACS-CS ensembles, we consider four random spatial source locations while on
the lightest ensemble we consider 32 random source locations.  These multiple samples are binned
and averaged as a single configuration estimate in the error analysis. Considering random spatial
sources, their cycle in time, the number of gauge fields and the consideration of 5 magnetic
fields, this results in 250 K unique correlation functions for each baryon at the lightest PACS-CS
quark mass.

% 32 * 8 * 197 * 5 = 250K

\section{Exceptional Configurations}\label{sec:exceptionalconfigurations}

\begin{figure*}
    \includegraphics[width=0.457\linewidth]{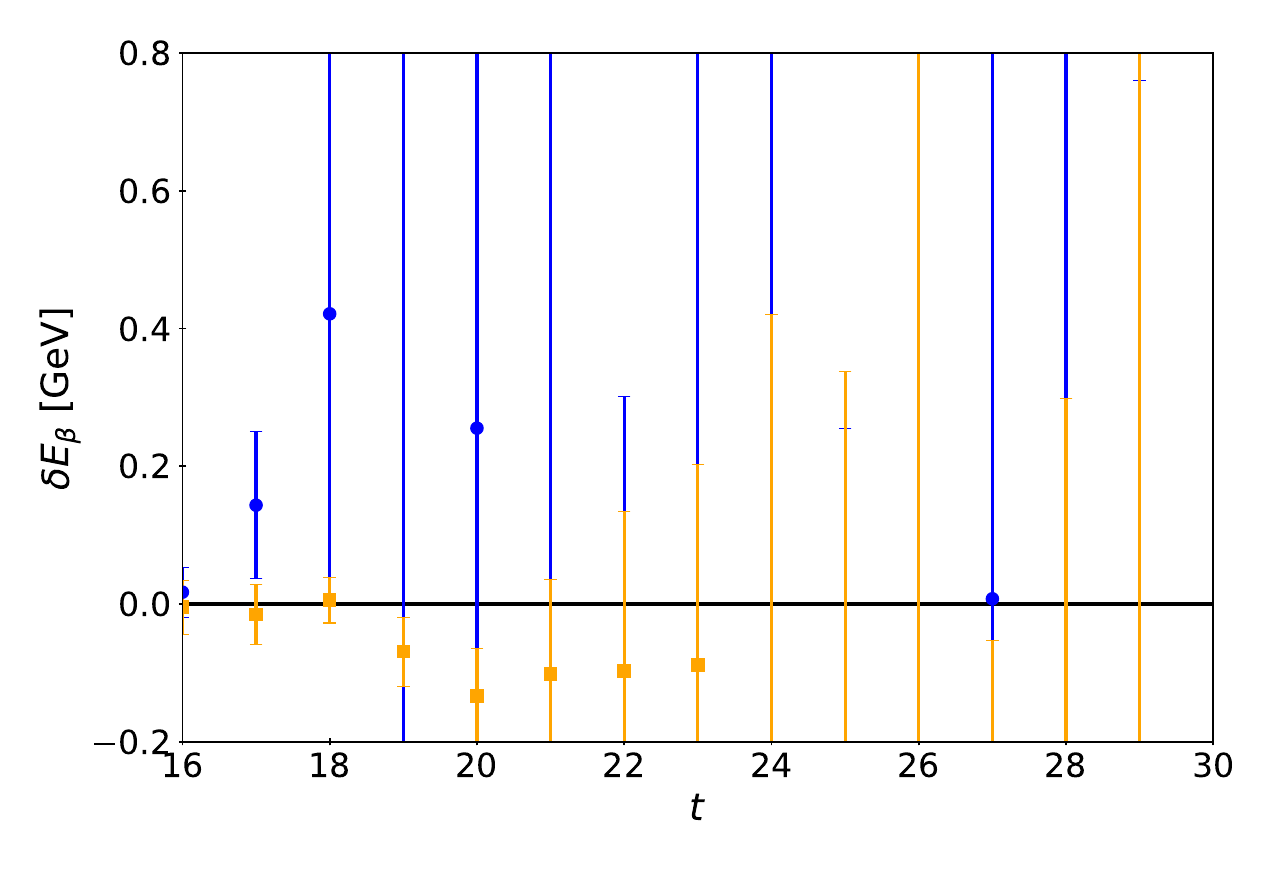}
    \quad
    \includegraphics[width=0.45\linewidth]{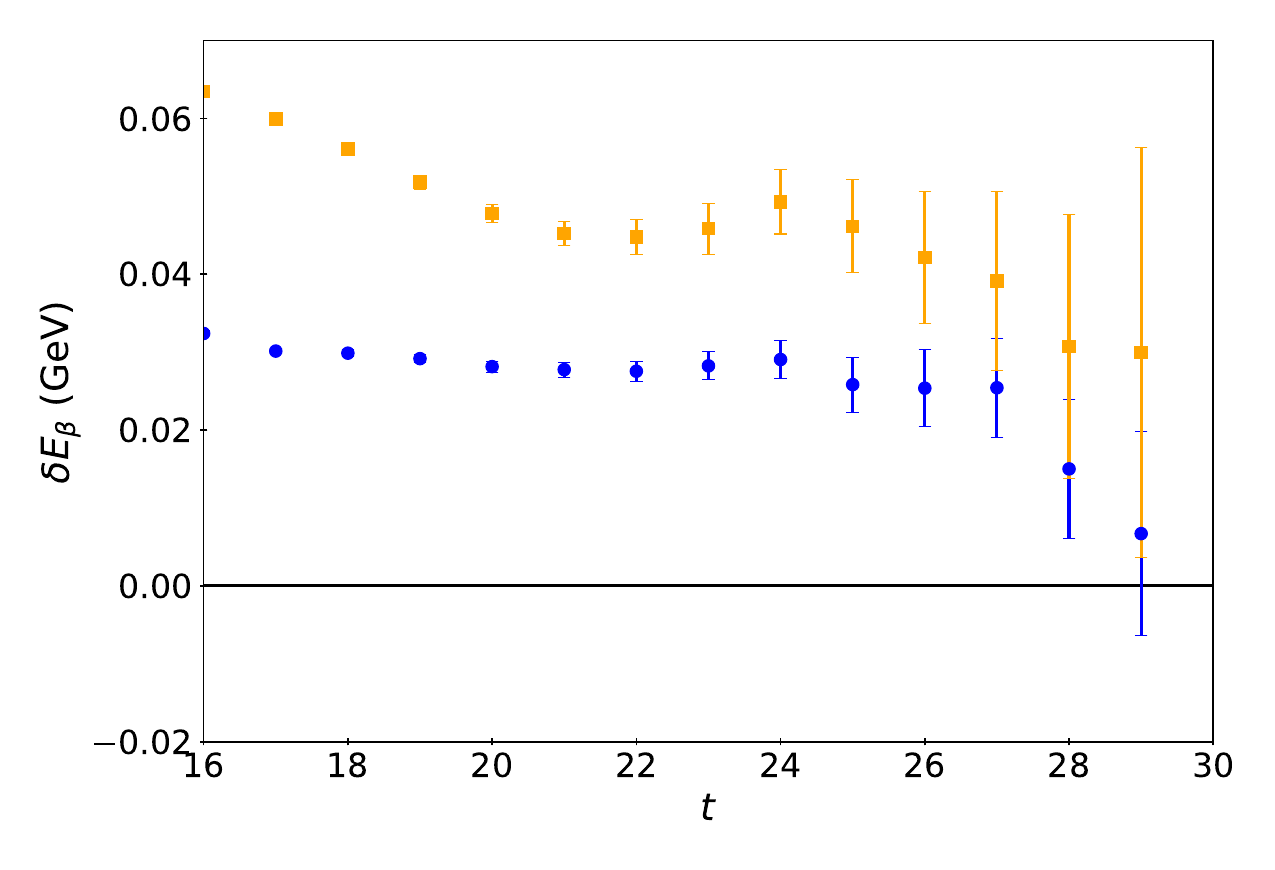}
    \caption{The magnetic polarisability energy shift $\delta E_{\beta}$ of
        \autoref{eqn:polarisability:energy:ratio} for a proton on the lightest (left) and second
        lightest (right) PACS-CS ensembles labelled by $m_\pi = 156$ and $296$ MeV
        respectively. Noting the $y$-axis limits, we see clear signs of exceptional behaviour with a
        wildly fluctuating plateau at the lightest mass.  This is in spite of a four-fold increase
        in statistics.
        % (32 * 197) / (4 * 399) = 4
    }
    \label{fig:exceptionalconfigurations:plateaucomparison}
\end{figure*}

Looking at the full untouched set of configurations on the lightest PACS-CS gauge ensemble, we see
clear signs of unusual behaviour. \autoref{fig:exceptionalconfigurations:plateaucomparison} shows
the difference in plateau behaviour for the magnetic polarisability energy shift $\delta E_{\beta}$
of \autoref{eqn:polarisability:energy:ratio} for the proton on the lightest PACS-CS ensemble (left)
and second lightest PACS-CS ensemble (right) labelled by $m_\pi = 156$ and $296$ MeV
respectively. This highly erratic behaviour of the magnetic polarisability energy shift does not
respond to increased statistics. As we will demonstrate, the cause of this behaviour is exceptional
configurations.

In this section, we begin by reviewing the origin of exceptional configurations and then describe
the algorithm we have created to systematically remove them.

\subsection{Wilson-type fermions}

The computationally efficient Wilson-clover fermion action used in this work breaks global chiral symmetry,
introducing an additive mass renormalisation to the Dirac operator. This shifts the point where the
renormalised quark mass is zero away from the point where the bare tree-level quark mass
vanishes. Of course this effect is well known and the solution is to consider the pion mass as a
function of the bare quark mass.  One need only determine the bare quark mass where the pion mass
vanishes (usually by linear extrapolation of $m_\pi^2$) and consider bare quark masses relative to
this point, referred to as the critical quark mass.  This principle is used in generating the 2+1
flavour dynamical configurations considered here.

Problems arise when improbable configurations are considered.  Historically, this problem was
encountered in the quenched approximation to QCD.  Gauge configuration probabilities that would be
suppressed by the fermion determinant in full QCD were admitted to the ensemble.  These improbable
configurations could induce uncharacteristic additive mass renormalisations that would lead to
divergent behaviour.

In the generation of dynamical-fermion gauge configurations the fermion determinant acts to
suppress the probability of creating configurations which give rise to approximate zeromodes of the
Dirac operator. In the quenched approximation, the fermion determinant is set to a constant and a
proliferation of approximate zeromodes is encountered. At sufficiently light renormalised quark
mass, it is possible to have the critical mass realised on a particular configuration near the bare
mass~\cite{Boinepalli:2004fz}, introducing a divergence in the quark propagator.  Such ``exceptional'' configurations
provide hadronic correlation functions which differ significantly in magnitude from the average,
introducing large statistical uncertainties and spoiling the ensemble average result.

In the early days of lattice QCD, when the calculation of a quark propagator would be the dominant
computational cost, it was common to only consider one fermion source per gauge-field
configuration.  Divergent behaviour was then attributed to the configuration.  However, in modern
simulations with several fermion sources per gauge field, it is now understood that the divergent
behaviour depends on both the configuration and the source position.  In the following, we preserve
the original terminology of an ``exceptional configuration,'' however it should be understood that
``configuration'' refers to a particular source on a particular gauge field.

The common consideration of electro-quenched configurations in the background field formalism
admits the opportunity to consider improbable gauge fields in the ensemble; these are gauge fields
that would be suppressed by the fermion determinant if the dynamical sea quarks were not blind to
the presence of the background magnetic field.  Again, additive mass renormalisation on an
improbable gauge field can induce divergent behaviour when the bare quark mass approaches the
critical quark mass, as it does for the lightest PACS-CS ensemble.

Exceptional configurations manifest in this calculation in two ways. Rarely, the conjugate-gradient
algorithm will fail to invert the Dirac operator at the required tolerance due to the singular
nature of the propagator. More commonly, a fermion-matrix inversion converges providing a quark
propagator with uncharacteristic behaviour. As an outlier, it spoils the ensemble average and
contributes to create a drastically increased uncertainty that not does respond to increased
statistics.

We observe that each baryon responds differently to exceptional configurations, with the
sensitivity governed by the role the propagator plays in the correlation function and the
environment that it is within.  We find that the inclusion of strange quarks increases the
tolerance of baryon correlation functions to exceptional configurations.  Nevertheless, the
presence of light quarks remains problematic.  Our worst case is that of the proton, where the
doubly represented $u$ quark requires the largest magnetic field strength with $k_d=\pm 4$.  Still
the rate at which exceptional configurations are identified in the following, even for this
worst-case example, is only 1.4\%, far below the quenched occurrence of 6\% for $m_\pi = 306(7)$
MeV reported in Ref.~\cite{Boinepalli:2004fz} using the favourable mean-field improved fat-link
irrelevant clover (FLIC) fermion action \cite{Zanotti:2001yb}.

\subsection{Identifying exceptional configurations}\label{sec:exceptionalconfigurations:algorithm}

Recalling that an exceptional configuration depends on both the configuration and the fermion
source position, our full list of $N$ ``configurations'' to examine is the product of the 197 gauge
field configurations in the lightest PACS-CS ensemble and the full set of randomised fermion source
locations.
We utilise the baryon correlation function determined on small sets of these configurations as a
proxy to check each configuration for exceptional behaviour.  Each baryon is analysed for its
unique set of exceptional configurations.

We consider a full set of $N$ correlation functions in random order and examine a subset of $M$
correlation functions. To determine the effective energy and associated uncertainty we:
\begin{enumerate}
    \item Construct a periodic list of correlation functions of length $N+M-1$.
    \item Given the $i$th subset of correlation functions $\{C_i, ..., C_{i+M-1}\}$ determine the effective energy $E_i(t)$ and uncertainty $\Delta E_i(t)$.
    \item Repeat for $i \in [1,N]$.
\end{enumerate}
We emphasise that the $i$th elements of $\{E_1,\cdots E_{N}\}$ and $\{\Delta E_1,\cdots \Delta
    E_{N}\}$ are the effective energy and associated uncertainty of the corresponding subset of
correlation functions in the $i$th subset. We'll refer to evolution of the index $i$ as evolution
in configuration time.
It is also important to emphasise that these quantities are distinct from the magnetic
polarisability alone as we seek a method of identifying exceptional configurations which is not
tied to the final desired observable.

\begin{figure}
    \includegraphics[width=\linewidth]{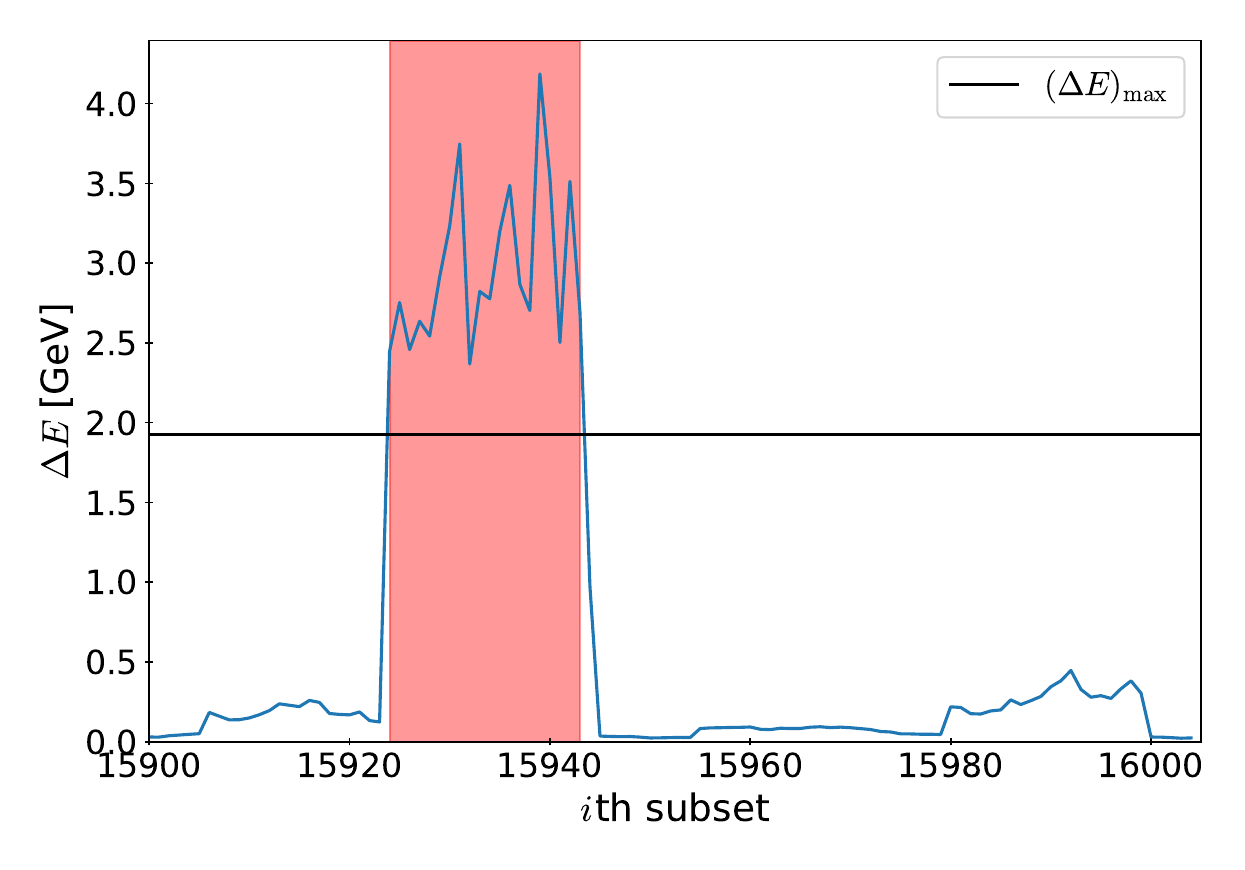}
    \caption{The step function behaviour of $\Delta E_i(t=18)$ is illustrated as a function of
        configuration time for the $i$'th subset. The jump in
        the uncertainty is caused by the inclusion of an outlying exceptional configuration
        contribution to the average over $M=20$ configurations. The horizontal line shows the
        exceptional configuration identification threshold which is updated after each iteration of
        the algorithm as described in the text. Another exceptional configuration of smaller
        magnitude beginning at $i=15980$ will be identified in a subsequent iteration.}
    \label{fig:exceptionalconfigurations:stepbehaviour}
\end{figure}

Figure \ref{fig:exceptionalconfigurations:stepbehaviour} presents a plot of the
uncertainty of the effective energy, $\Delta E_i(t)$, for the $i$'th configuration subset of
$M = 20$ configurations at $t=18$, two time slices following the source at $t=16$.
We have found it is
helpful to search for exceptional configurations at early time slices where the statistical fluctuations
are normally small.
Plotting these quantities as a function of configuration time results in an
approximately constant line which exhibits sharp step function behaviour whenever an exceptional
configuration enters the subset of $M$ considered configurations.  The step lasts for precisely
$M$ steps in configuration time until the exceptional configuration leaves the subset.

The identification of an exceptional configurations requires identification of when these sharp
changes occur. While the sub-ensemble average also sees a step change, we find the step change
in the uncertainty in the effective energy to provide a more robust indicator with greater
sensitivity.

\begin{figure}
    \includegraphics[width=\linewidth]{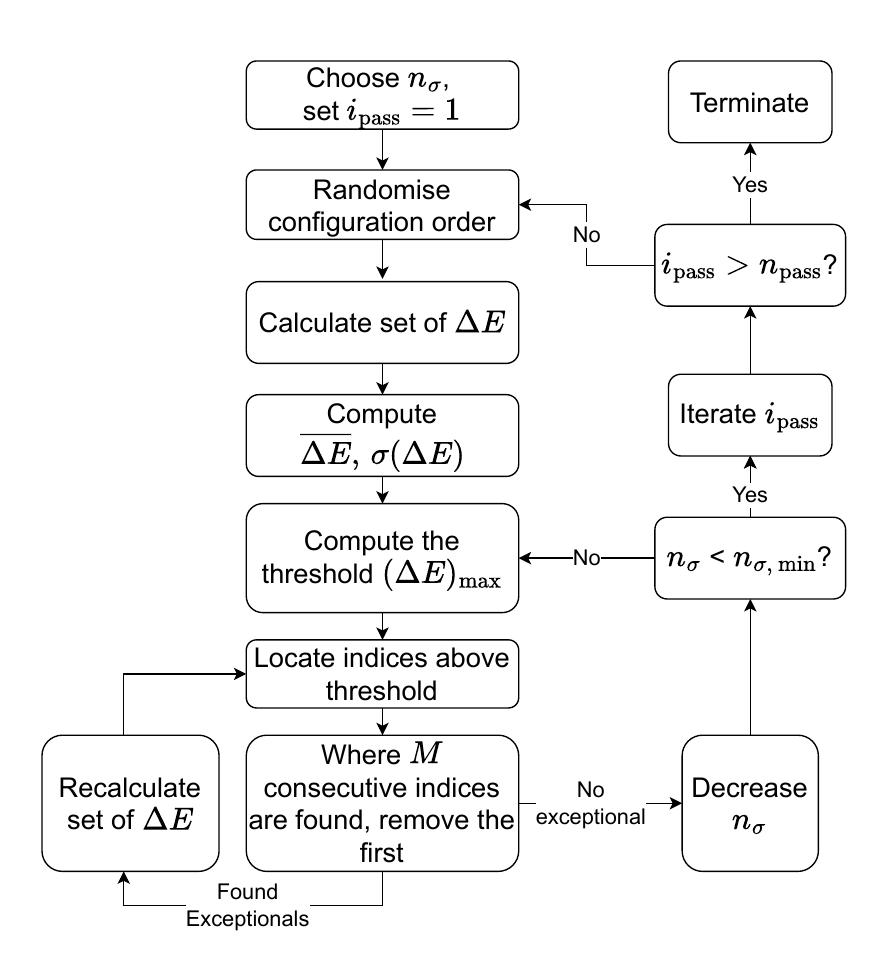}
    \caption{Flowchart of the exceptional configuration identification algorithm described in the text.
    }\label{fig:exceptional:flowchart}
\end{figure}

Our algorithm, presented in detail in the following, is presented visually in
\autoref{fig:exceptional:flowchart}.
The algorithm runs concurrently across a range of time slices, from $t=18$ two time slices beyond
the source to where we expect to lose signal, $t = 25$. In practice, most exceptional
configurations are identified at time slices close to the source as the rapidly increasing noise
associated with the baryon correlation function obscures exceptional configurations at later time
slices.  If a configuration is found to be exceptional on any time slice, it is removed from the
configuration list.
Understanding that the calculations are performed concurrently at all relevant time slices, we
drop the time dependence in the following discussion.

We commence by determining a threshold (corresponding to the horizontal line in
\autoref{fig:exceptionalconfigurations:stepbehaviour}) above which a configuration subset's energy
uncertainty is considered exceptional.  We consider the entire set of sub-ensembles with uncertainties
$\{\Delta E_1,\cdots \Delta E_{N}\}$ and calculate the mean $\overline{\Delta E}$ and standard
deviation of the distribution of uncertainties, $\sigma(\Delta E)$.
The threshold is set as
\begin{equation} \label{eqn:exceptional:threshold}
    (\Delta E)_{max} = \overline{\Delta E} + n_{\sigma}\,\sigma(\Delta E) \, ,
\end{equation}
where $n_{\sigma}$ is the number of standard deviations from the mean. We typically begin with
$n_{\sigma}=20$, a large conservative value to ensure that no exceptional configurations are found
initially.

Using the threshold calculated above, we construct a list of all $i$ sub-ensembles where $\Delta
    E_i > (\Delta E)_{max}$ and identify where we have at least $M$ consecutive indices. In this
case, the first of the consecutive indices corresponds to an exceptional configuration. We allow
the number of consecutive indices to be greater than $M$ as this allows for the removal of
exceptional configurations which superpose other configurations due to being close to each other in
the list.

Identified exceptional configurations are removed, triggering a recalculation of the $E_i$ and
$\Delta E_i$.  However, we do not update $(\Delta E)_{max}$ as subsequent iterations at this point
are aimed at removing exceptional configurations that overlapped within the sub-ensembles of $M$
configurations and therefore were not eliminated. A new list of all $i$ sub-ensembles where $\Delta
    E_i > (\Delta E)_{max}$ is created.  Again, where there are at least $M$ consecutive indices the
first index is identified as exceptional.

This process of removal, updating $E_i$ and $\Delta E_i$ and identification of exceptional configurations
continues until no new configurations are identified as exceptional.  This process ensures the
algorithm is as conservative as possible.

Whenever exceptional configurations are unable to be found at the established threshold,
we then decrease $n_{\sigma}$ by a small amount and recalculate $(\Delta E)_{max}$.
In practice, we decrease $n_{\sigma}$ by 0.1 at each step.

Our aim is to save a configuration list surviving at a target threshold $n_{\sigma,\,{\rm min}}$.
We will run the algorithm over values of $n_{\sigma,\,{\rm min}}$ from 20 down to 2 with a target
step size of 0.25. The fine grain and broad range allow for very precise identification of when the
character of the effective energy shift plateau changes.  We note that each target value starts
from a different randomised ordering of the configurations and this further aids in identifying the
best target value for $n_{\sigma,\,{\rm min}}$ for eliminating the minimum number of exceptional
configurations possible.

Thus, the algorithm continues to iterate until $n_{\sigma,\,{\rm min}}$ is reached.  At the
termination of the algorithm we save the list of configurations that have survived the criteria for
the further analysis. This constitutes one pass of the algorithm.

In addition, there is some dependence in the identification of exceptional configurations on the
ordering of the configurations.  Upon completion of a pass, the list of remaining configurations is
randomised. The value of $n_\sigma$ is reset to the initial value, and the algorithm runs again.
This repeats until a predetermined number of passes is completed.  Here we create configuration
lists for $n_{\rm pass} \in \{2,3,4,5,6\}$.

The configuration lists for specific values of $n_{\sigma,\,{\rm min}}$ and $n_{\rm pass}$ are then
sorted by the number of configurations surviving.  This enables an examination of the behaviour of
the magnetic polarisability energy shift profile as a function of the number of exceptional
configurations removed for each baryon separately.  Animations are provided in the supplementary
information.  We find these animations to be particularly helpful in determining the minimum number
of configurations to eliminate to obtain a step change in the statistical errors of the
energy shift.

\subsection{Field strength dependence of exceptional configurations}

It is the inclusion of the background magnetic field which causes the presence of exceptional
configurations due to the electro-quenching associated with neglecting the background magnetic field
in the gauge-field generation process. As a result, larger magnetic field strengths produce greater
rates of exceptional configurations.
As a result, we examine the largest baryon magnitude field strengths $k_d=\pm 2$ first.  For
baryons with a $u$ quark, field strengths with $k_d=\pm 4$ are included in the calculation of the
baryon correlation function.

In determining the magnetic polarisability, we consider baryon spin-field aligned and
anti-aligned orientations for each of the magnetic field strengths. Extensive testing was done to
determine the best approach for considering these spin-field orientations.  We find exceptional
configurations are most efficiently identified when the spin-field aligned and anti-aligned
correlation functions are averaged for a given magnetic-field orientation, effectively averaging
spin up and down baryon orientations.

Thus, baryon field strengths $k_d=2$ and $k_d=-2$ are considered separately and we take the union
of the set of exceptional configurations found for each field. We then progress to checking
$k_d=\pm 1$ separately having already removed the configurations found at $k_d=\pm 2$. Checking the
positive and negative field strengths separately ensures that neither field strength is
unnecessarily biased. Checking the larger magnitude field strength first ensures we omit the
majority of configurations which are exceptional as early as possible.

\subsection{Summary}

We emphasise that it is unnecessary to explicitly check the $k_d=0$ correlation functions, as we do
not anticipate the presence of any exceptional configurations in the zero-field case. A very small
number of ``exceptional configurations'' can be found with our algorithm, but their removal has no
impact on the profile of the effective energy.

As we will see in the following section reporting our results, it is not easy to define a
consistent stopping criterion for when to cease removing exceptional configurations identified by
our algorithm. Moreover, each of the octet baryons have different susceptibilities to exceptional
configurations and require different proportions of exceptional configurations to be removed.
It is here that the animations of the supplementary information are particularly helpful.

\section{Lattice Results}\label{sec:latticeresults}

\begin{table}[tb]
    \centering
    \caption{Baryon magnetic polarisability values extracted through fitting the magnetic
    polarisability energy shift $\delta E_{\beta}$ of Eqs.~\ref{eqn:polarisability:energy:ratio}
    and \ref{eqn:polarisability:energyshift} over the range $[t_{\rm min}=20, t_{\rm max}]$. In
    the case of the hyperons, there is sufficient signal available without removing any
    exceptional configurations to allow fitting. These values are included in the ``Untouched''
    column. The ``Exceptional removed'' column are the results from fitting to the rightmost
    effective energies in \autoref{fig:results:nucleonplateaus},
    \autoref{fig:results:sigmaplateaus}, and \autoref{fig:results:cascadeplateaus}.
    The zero-field fit range is $[t_{\rm min},t_{\rm max}] = [20, 30]$.
    Magnetic polarisability values are in units $\times
        10^{-4}\,$fm$^3$.
    } \label{tab:latticeresults:polarisabilityvalues}
    \begin{ruledtabular}
        \begin{tabular}{ccccc}
            \noalign{\smallskip}
                      & \multicolumn{2}{c}{Untouched} & \multicolumn{2}{c}{Exceptional removed}                            \\
                      & $t_{\rm max}$                 & $\beta$                                 & $t_{\rm max}$ & $\beta$  \\
            \noalign{\smallskip}\hline\noalign{\smallskip}
            $p$       &                               &                                         & 25            & 2.8(16)  \\
            $n$       &                               &                                         & 22            & 1.7(12)  \\
            \sigmap   & 22                            & 1.7(10)                                 & 26            & 2.17(51) \\
            \sigmam   & 25                            & 0.42(50)                                & 25            & 0.41(23) \\
            \cascadez & 25                            & 2.14(32)                                & 29            & 2.32(20) \\
            \cascadem & 24                            & 0.37(25)                                & 28            & 0.20(18) \\
            \noalign{\smallskip}
        \end{tabular}
    \end{ruledtabular}
\end{table}

Here we present the results for each baryon. We show the effective energy shift, $\delta
    E_{\beta}(B,t)$, of \autoref{eqn:polarisability:energy:ratio} for both baryon field strengths $k_d
    = 1$ and $2$.  We show the energy shift for the original case of no configurations removed, and
subsequently the results from the removal of the minimal number of exceptional configurations to
resolve the effective energy at $k_d = 1$ and $2$ respectively.

The determination of this minimal number of exceptional configurations follows from the animations
of the supplementary material, \autoref{fig:protonanimation} through \autoref{fig:cascademanimation}, showing the evolution of the effective energy
shift as the number of exceptional configurations removed increases.
The animations show that the quality of the effective energy shift does not gradually improve as
configurations are removed, rather the uncertainties reduce significantly and suddenly and resolve
the presence of plateau behaviour.
Due to the differing number of exceptional configurations observed at each field strength, the
$k_d=1$ plateau improves first, followed by the $k_d=2$ plateau. These are the results presented in
the following figures.

In \autoref{fig:results:nucleonplateaus} for the nucleons, \autoref{fig:results:sigmaplateaus} for
the $\Sigma$ baryons, and \autoref{fig:results:cascadeplateaus} for the $\Xi$ baryons we show the
effective energy shift in the initial state with no exceptional configurations removed on the left,
after the resolution of the $k_d=1$ plateau behaviour in the middle and after the resolution of the
$k_d=2$ plateau behaviour on the right. Note that the $k_d=2$ plateau shows some improvement as the
$k_d=1$ plateau is resolved, but further exceptional configuration removal to resolve the $k_d=2$
plateau on the right is crucial, typically extending the plateau further in Euclidean time.

In calculating the magnetic polarisability, we choose to fit the effective energy shift immediately
after the $k_d=2$ plateau resolves.
As discussed earlier, we utilise the weighted averaging method discussed in
\autoref{sec:background}, sampling fit windows beginning at $t_{\rm min}=20$ up to the point where
the signal is lost which varies for each baryon. The results and $t_{\rm max}$ values for all
baryons are summarised in \autoref{tab:latticeresults:polarisabilityvalues}.

\subsection{Nucleons}

\begin{figure*}
    \includegraphics[width=0.3\linewidth]{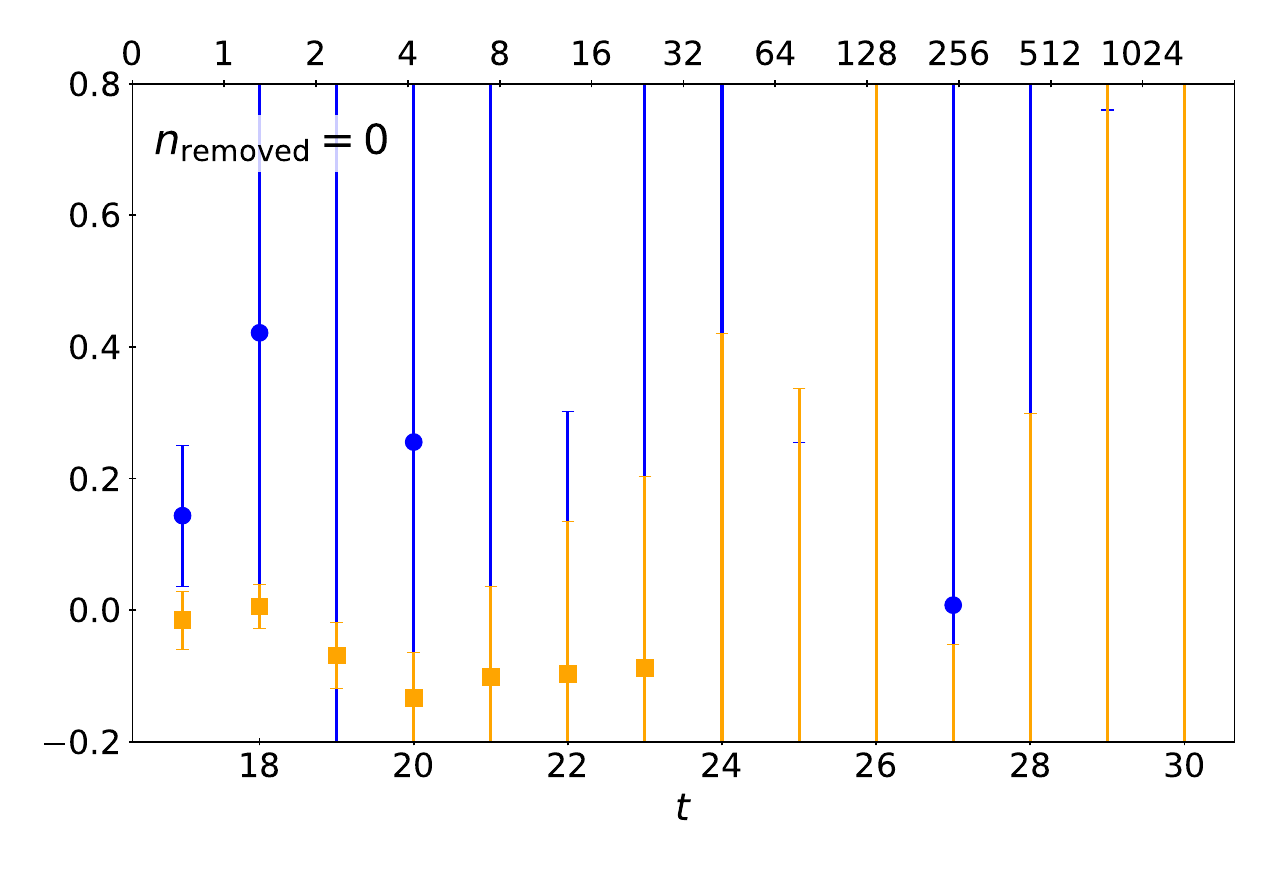}
    \includegraphics[width=0.3\linewidth]{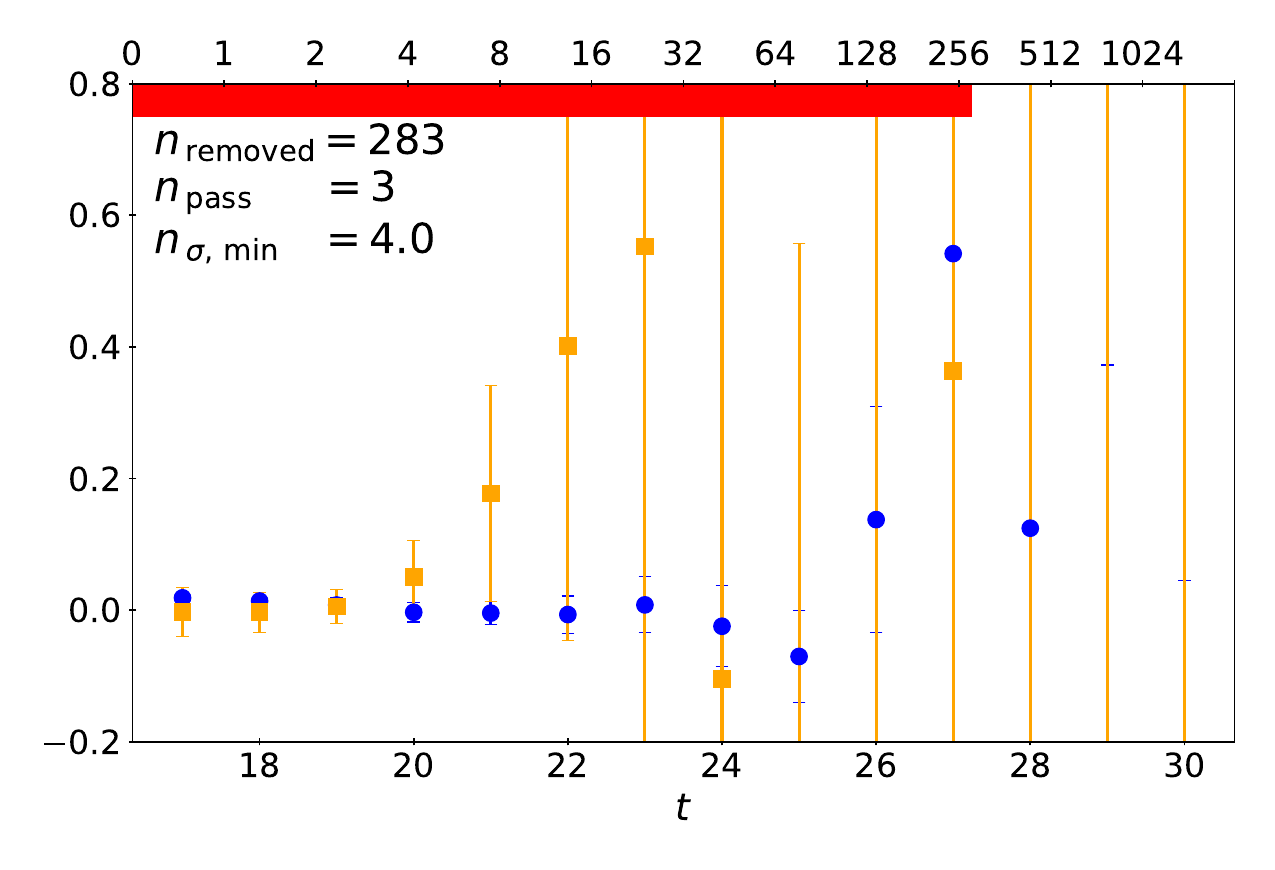}
    \includegraphics[width=0.3\linewidth]{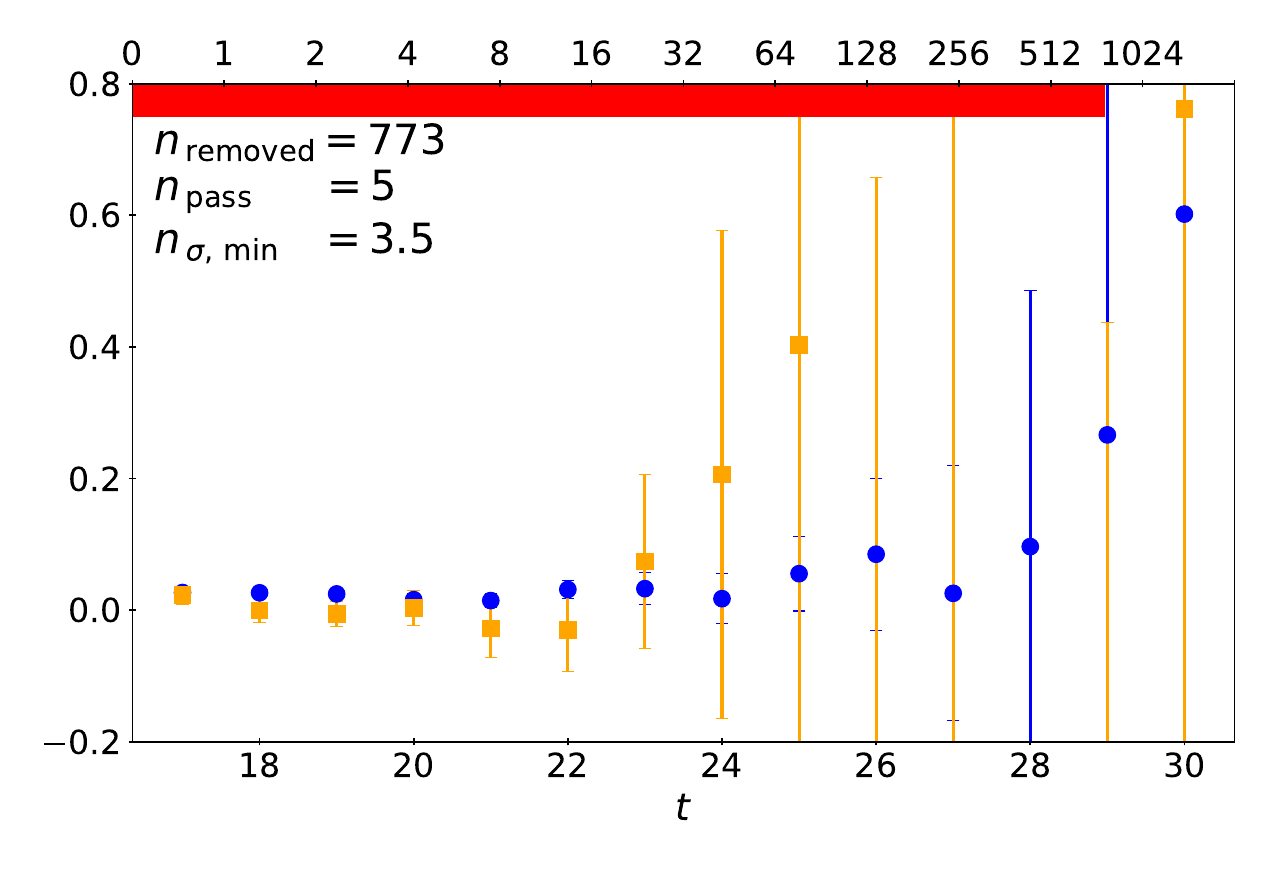}
    \includegraphics[width=0.3\linewidth]{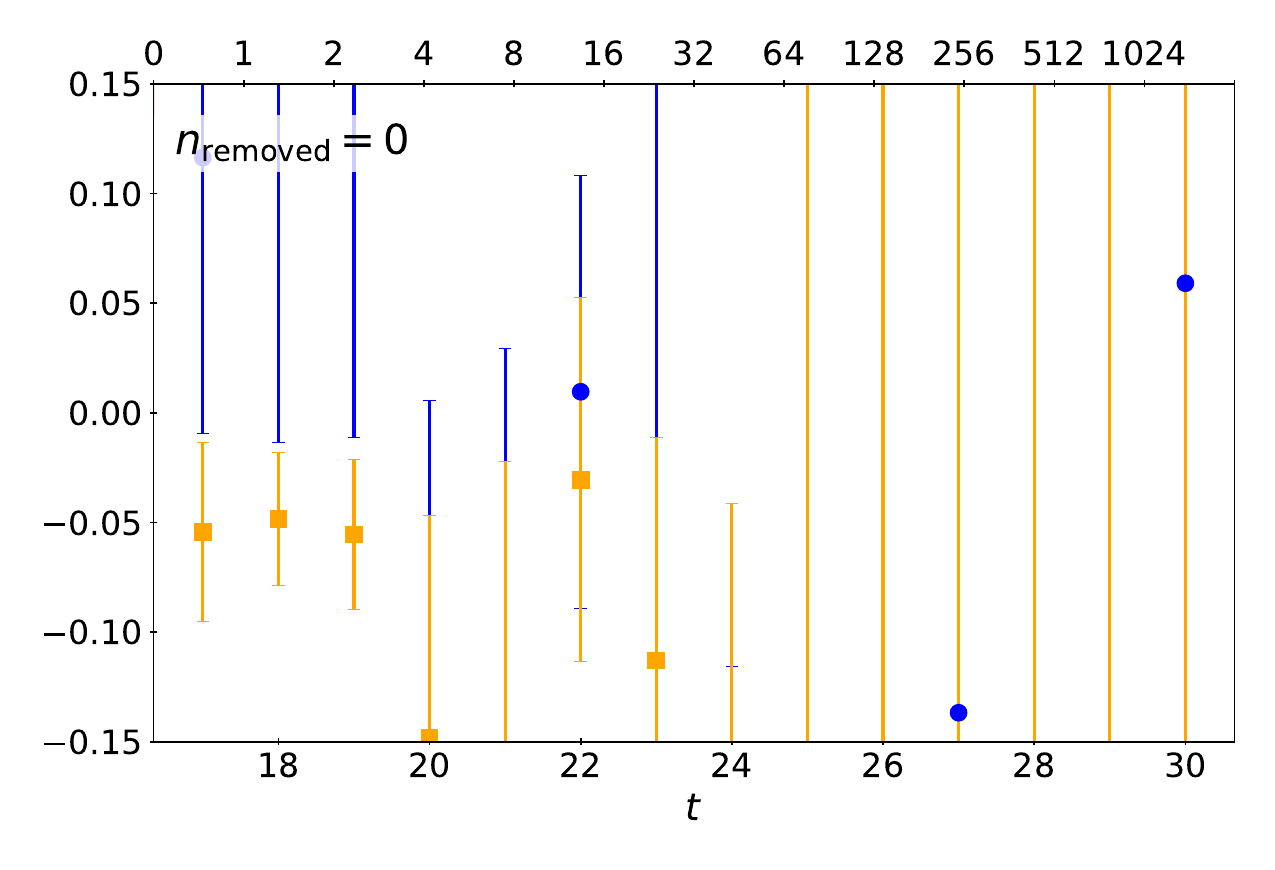}
    \includegraphics[width=0.3\linewidth]{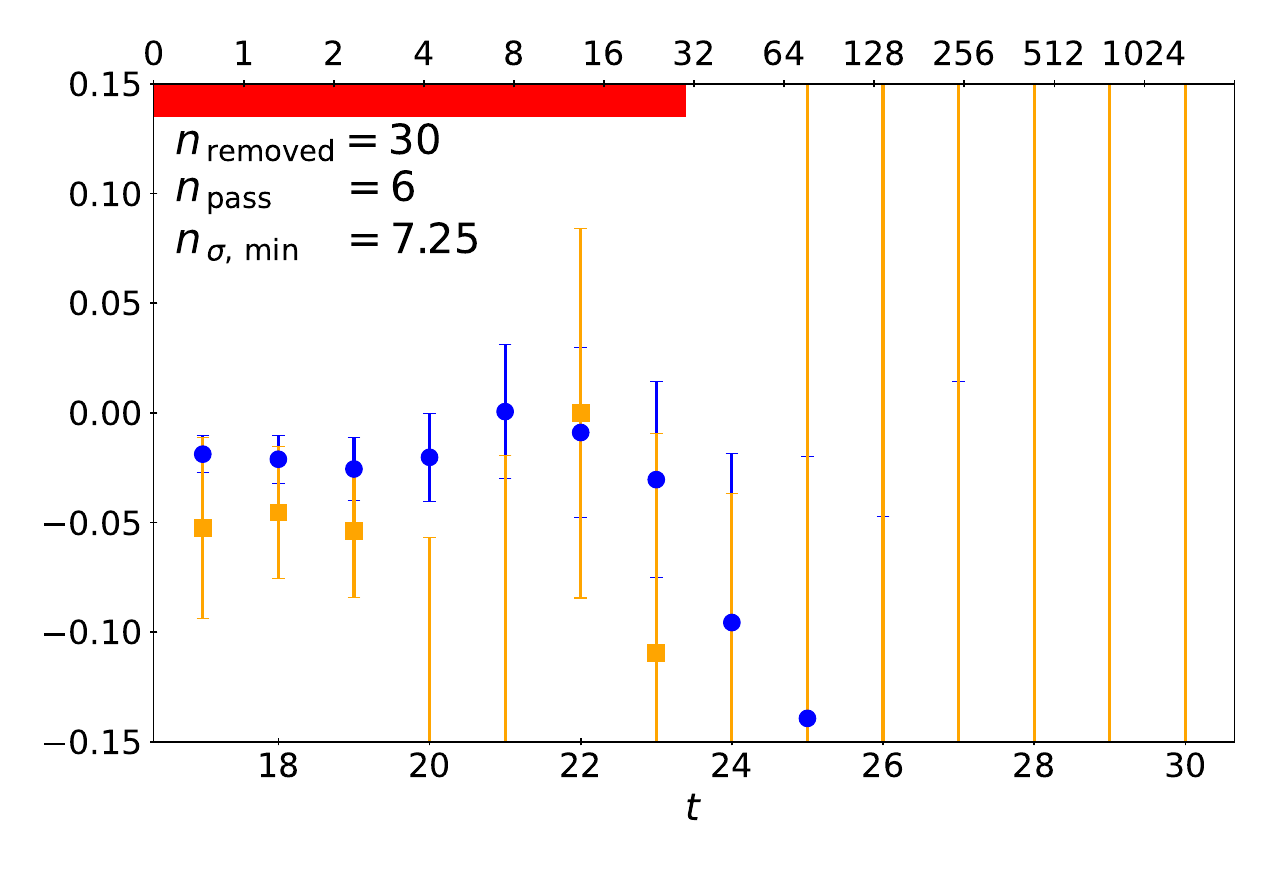}
    \includegraphics[width=0.3\linewidth]{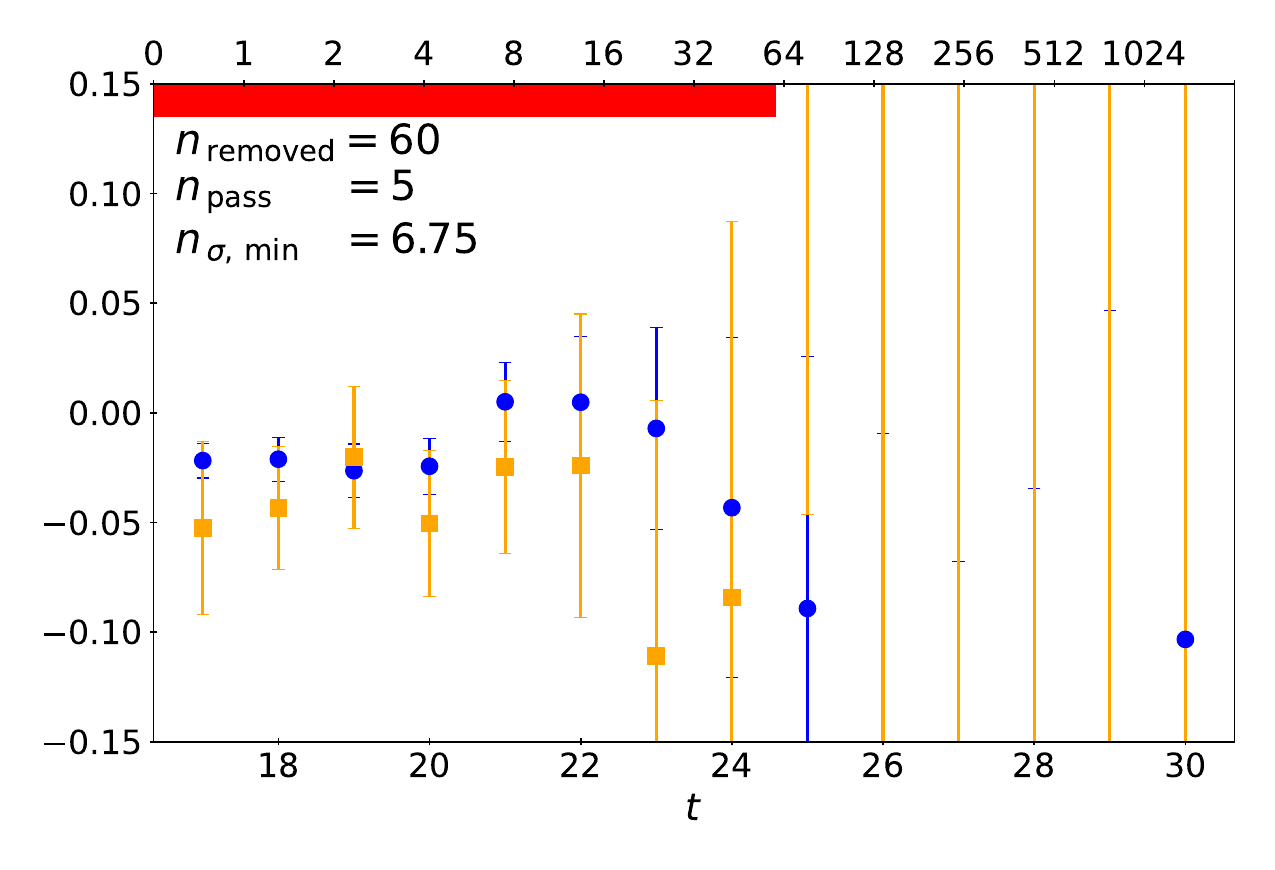}
    \caption{The effective energy shift $\delta E_{\beta}(B,t)$, of
        \autoref{eqn:polarisability:energy:ratio} for the proton (top) and neutron (bottom). On the
        left, there are no exceptional configurations removed, the $k_d=1$ field-strength plateau has
        just resolved in the middle plot and the $k_d=2$ plateau has just resolved in the right
        plot. The red bar at the top of the plots indicates the number of configurations removed to
        achieve the resolution presented.  This and the corresponding pass number and
        $n_{\sigma,\,{\rm min}}$ are also indicated.}
    \label{fig:results:nucleonplateaus}
\end{figure*}

The effective energy shift $\delta E_{\beta}(B,t)$, of \autoref{eqn:polarisability:energy:ratio}
for the nucleons are shown in \autoref{fig:results:nucleonplateaus}.
The proton is the most susceptible of the outer octet baryons to exceptional configurations
effects. This is due to the combination of three light quarks in its composition and the doubly
represented up quark which feels the background magnetic field more strongly than the down and
strange quarks. Due to the difficulty in extracting signal for the proton, we utilise the quality
of the $k_d=1$ plateau and fit in the range $[t_{\rm min},t_{\rm max}]=[20,25]$. In this case we
have removed 773 exceptional configurations of approximately 50 K configurations, a fraction of 1.4\%.

Containing three light quarks, the neutron is also affected strongly by exceptional configurations,
though the dynamics of the singular up quark reduce its susceptibility compared to the proton. This
can be seen in \autoref{fig:results:nucleonplateaus} by comparing the number of configurations that
need to be removed to achieve the resolution of the plateau at each field strength.

Due to the quickly decaying first field strength plateau, we do not attempt to push the fit later
in the case of the neutron, fitting in the range $[t_{\rm min},t_{\rm max}]=[20,22]$. The neutron
requires the removal of only 60 exceptional configurations. While we could leave the algorithm
running longer, we have already observed a jump in the quality of the plateau at both field
strengths. As such, continuing to remove exceptional configurations would risk removing
configurations that are valid signal. Indeed the magnetic polarisability is sensitive to continued
configuration removal, tending to increase to larger values. Thus it is important to remove the
minimal number of configurations required to resolve the effective energy plateau.

% Windows for easy reference
% proton:  [20,25]
% neutron: [20,22]

\subsection{Sigma baryons}

\begin{figure*}
    \includegraphics[width=0.3\linewidth]{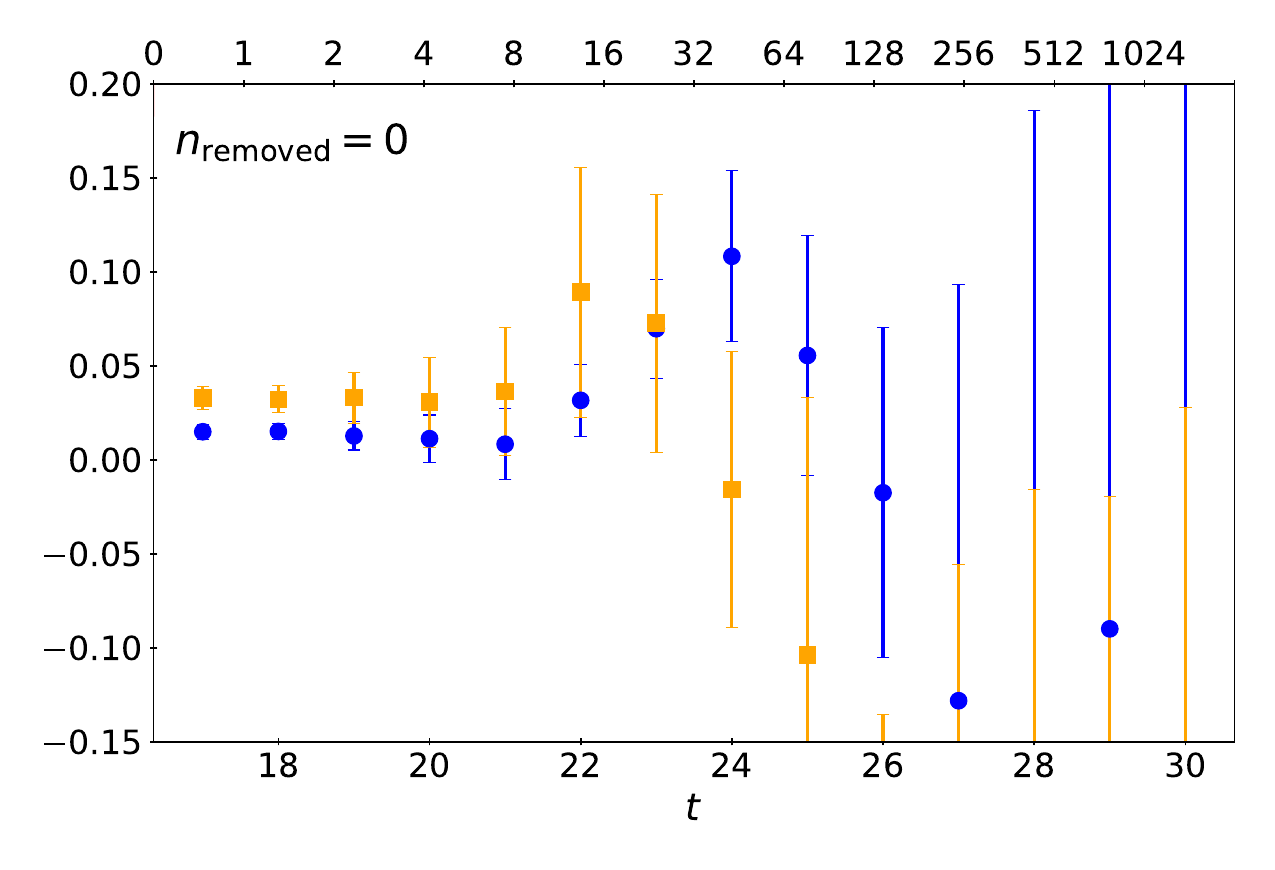}
    \includegraphics[width=0.3\linewidth]{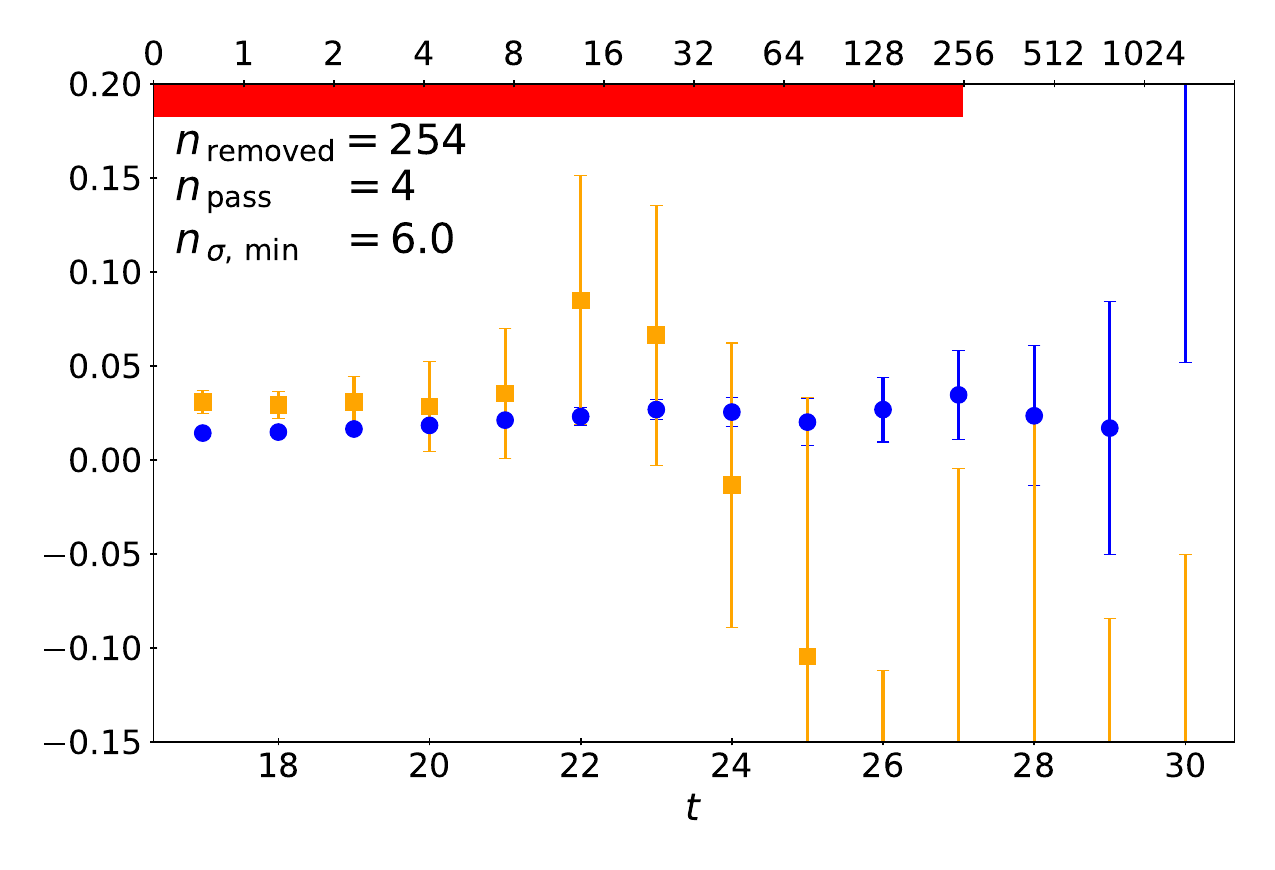}
    \includegraphics[width=0.3\linewidth]{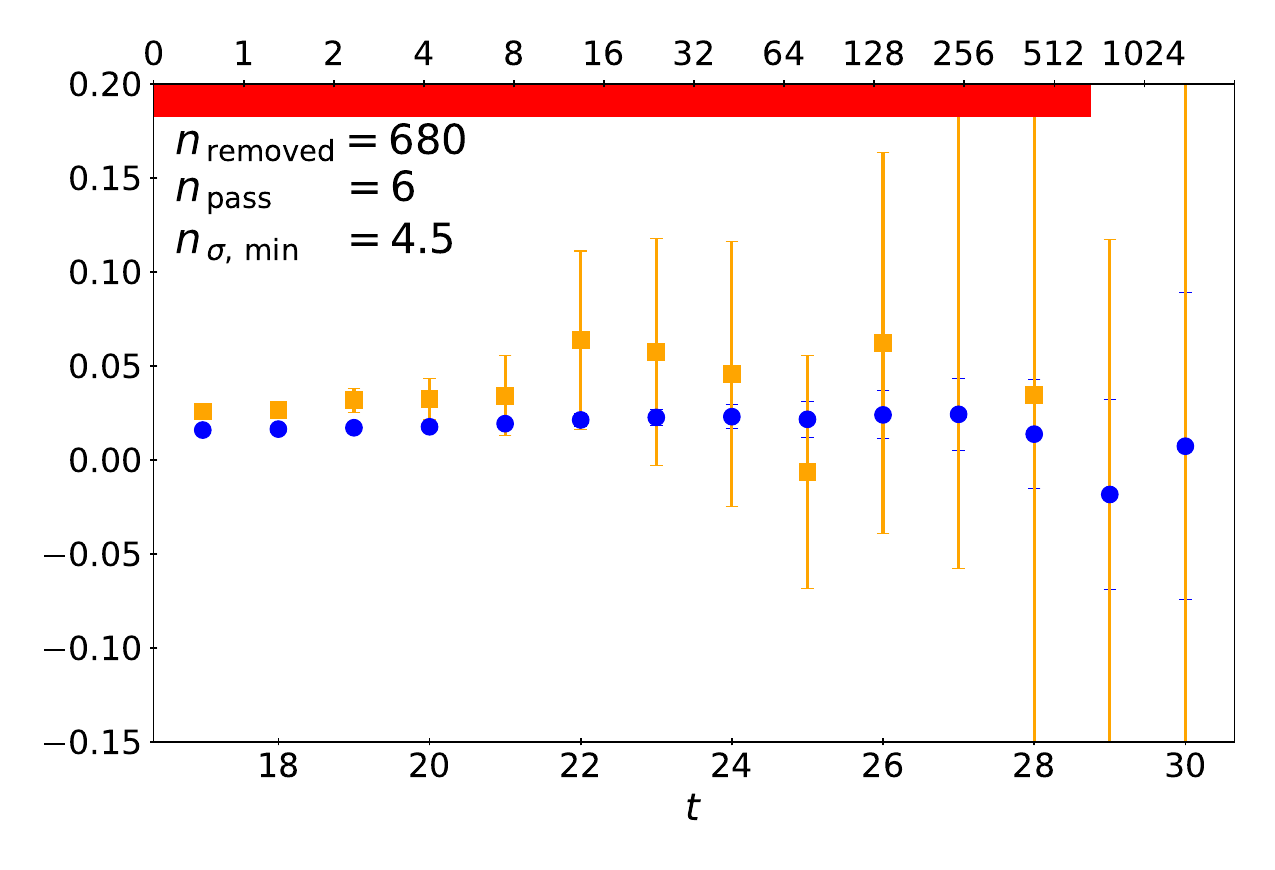}
    \includegraphics[width=0.3\linewidth]{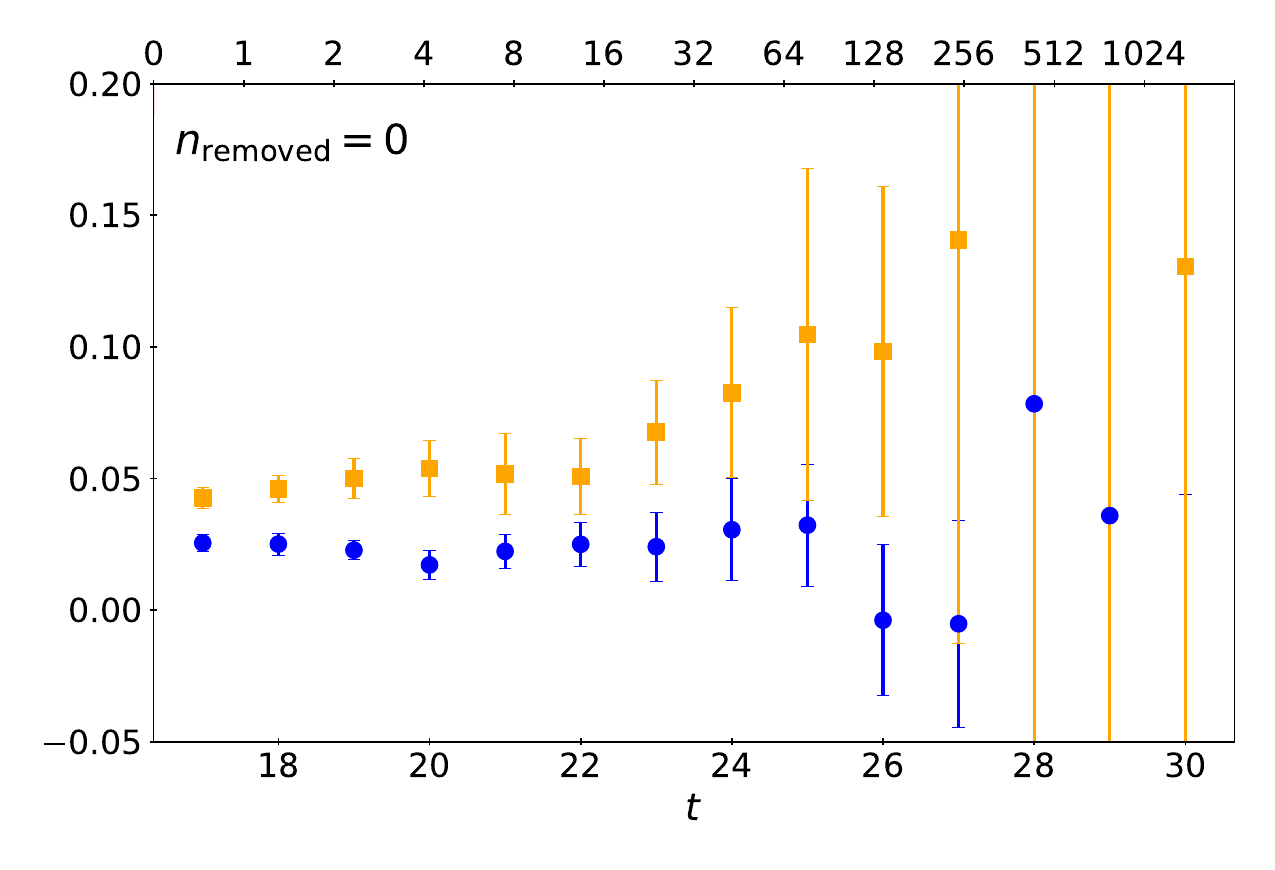}
    \includegraphics[width=0.3\linewidth]{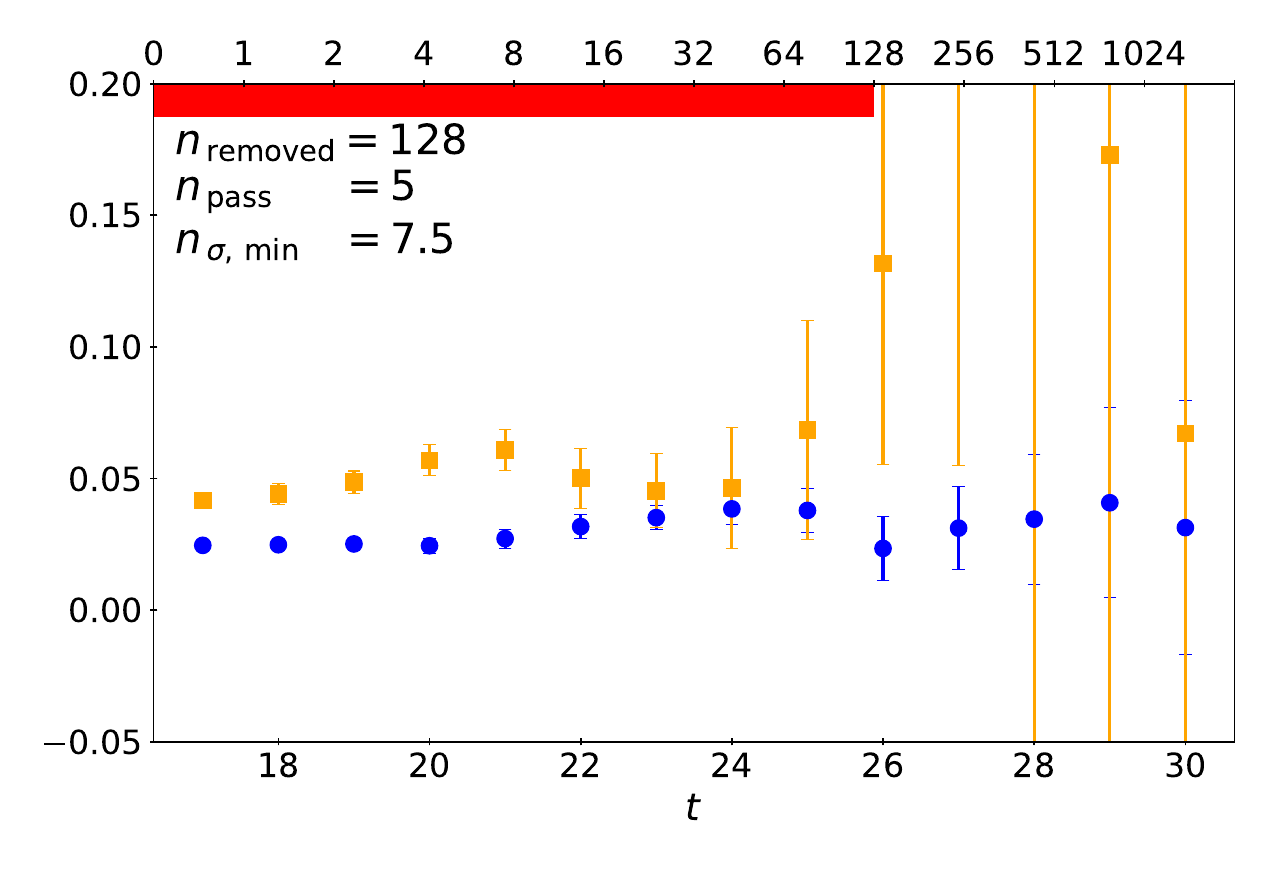}
    \includegraphics[width=0.3\linewidth]{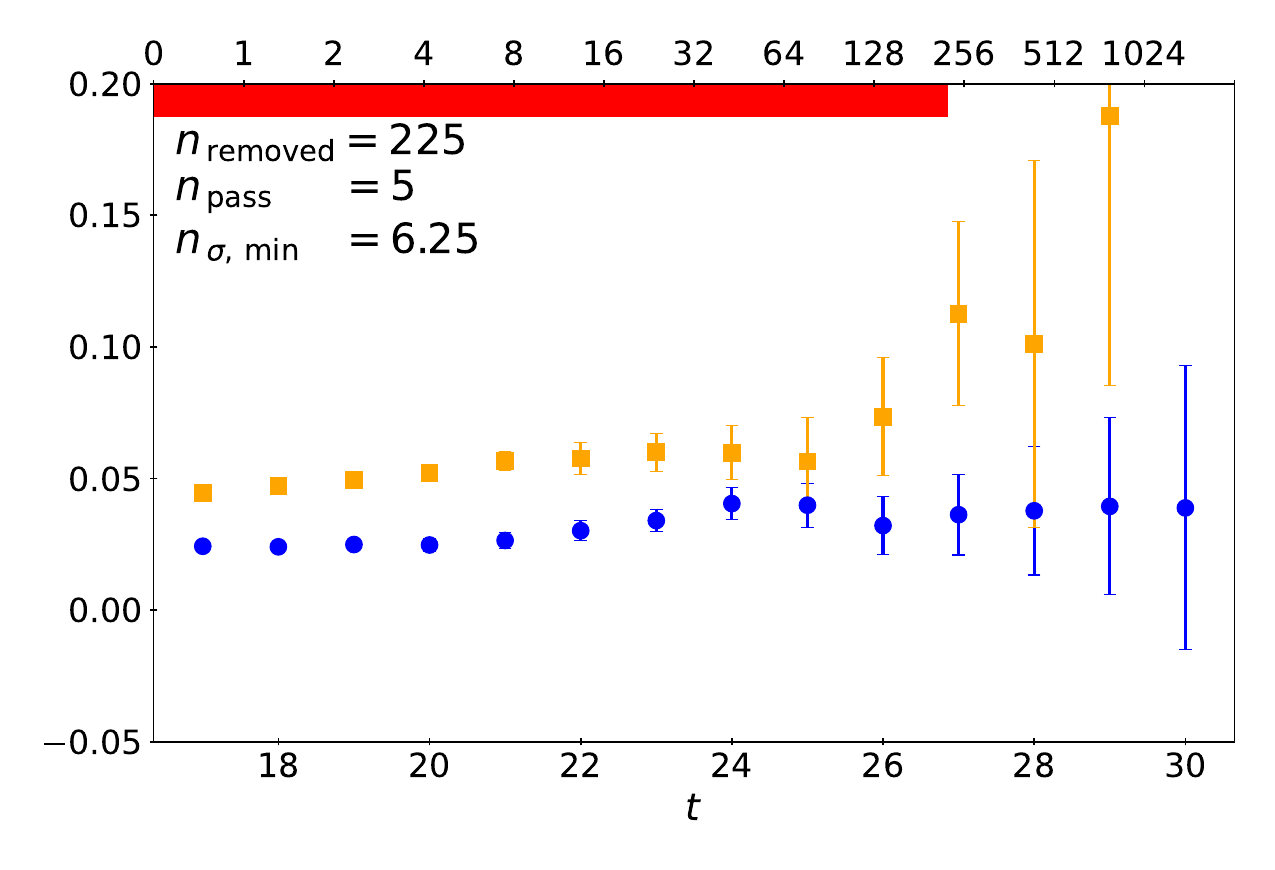}
    \caption{The effective energy shift $\delta E_{\beta}(B,t)$, of
        \autoref{eqn:polarisability:energy:ratio} for the \sigmap\ (top) and \sigmam\ (bottom).
        Plots and symbols are as in \autoref{fig:results:nucleonplateaus}.}
    \label{fig:results:sigmaplateaus}
\end{figure*}

The plateaus for the $\Sigma$ baryons are shown in \autoref{fig:results:sigmaplateaus}.
With the addition of a singular strange quark we see a significant reduction in the susceptibility
of the effective energy to exceptional configurations. To emphasise this reduced susceptibility,
for each of the hyperons, we will fit the complete untouched set of configurations in addition to
the plateaus improved through the removal of exceptional configurations.

As there is still a doubly represented up quark sector in the \sigmap, it is influenced more
strongly than the \sigmam\ in much the same capacity as the relationship between the proton and
neutron. This is demonstrated by the difference in number of exceptional configurations
removed. For the \sigmap\ we remove 680 configurations but only 225 for the \sigmam.

In the full ensemble case, we are only able to fit in the range $[t_{\rm min},t_{\rm max}]=[20,22]$
for the \sigmap\ due to its susceptibility while the absence of an up quark allows fitting of the
\sigmam\ in the range $[t_{\rm min},t_{\rm max}]=[20,25]$. With the removal of exceptional
configurations, we now see good plateau behaviour for the \sigmap\ and excellent plateau behaviour
for the \sigmam. The quality of the $k_d=1$ signal for the \sigmap\ allows fitting in the range
$[t_{\rm min},t_{\rm max}]=[20,26]$. Interestingly, the removal of exceptional configurations does
not allow for fitting longer for the \sigmam\ as the $k_d=2$ signal decays quickly beyond $t=25$,
though it does produce a more precise value.

% Windows for easy reference [untouched] [removed]
% sigmap: [20,22], [20,26]
% sigmam: [20,25], [20,25]

\subsection{Cascade baryons}

\begin{figure*}
    \includegraphics[width=0.3\linewidth]{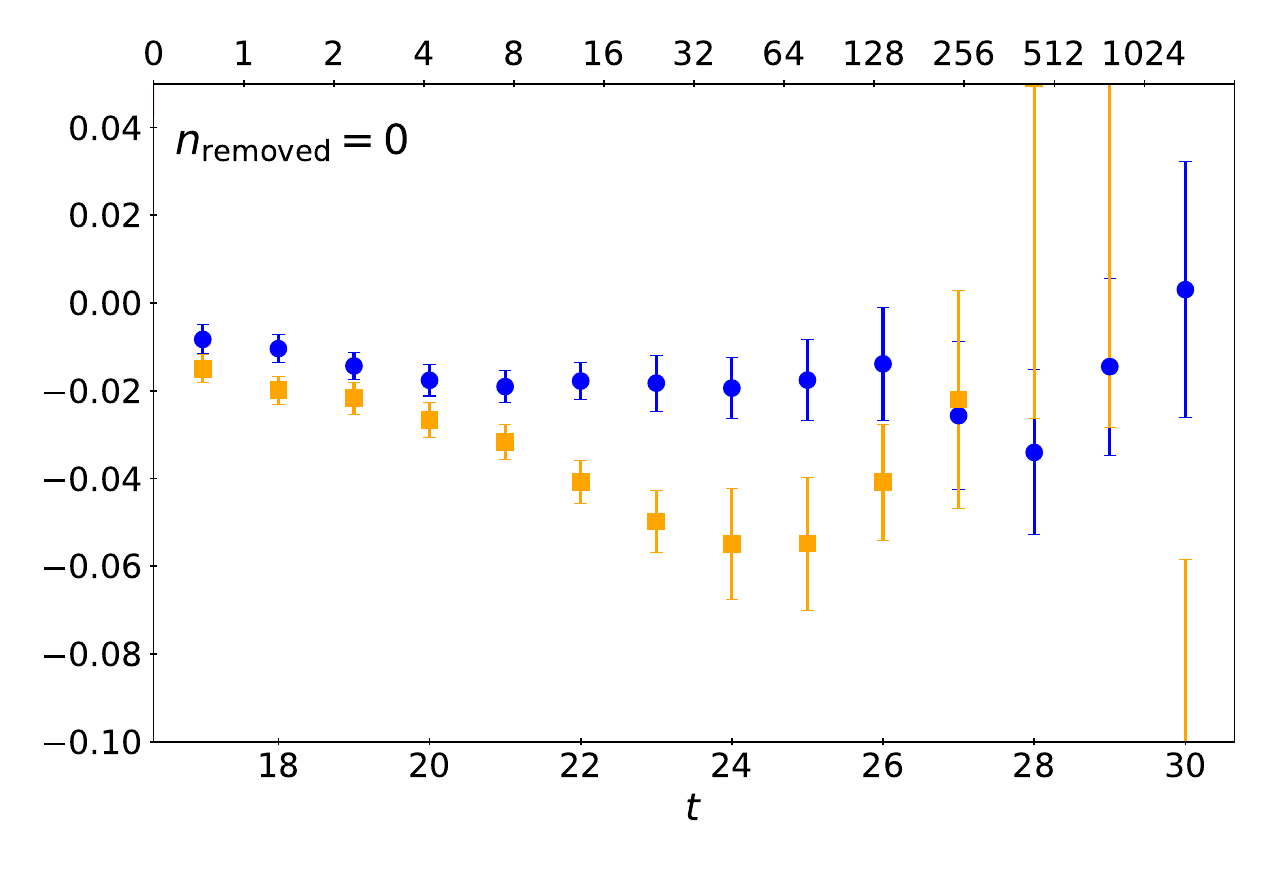}
    \includegraphics[width=0.3\linewidth]{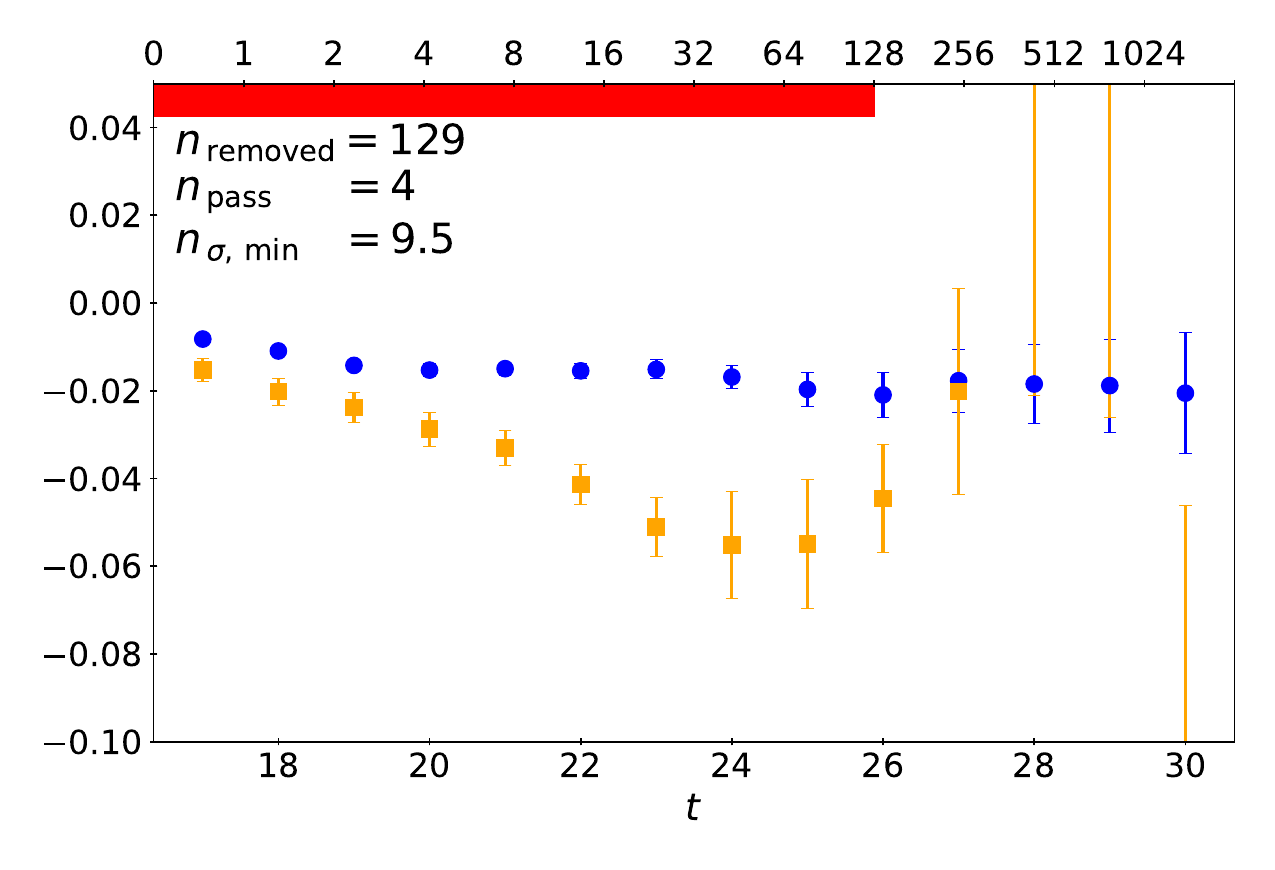}
    \includegraphics[width=0.3\linewidth]{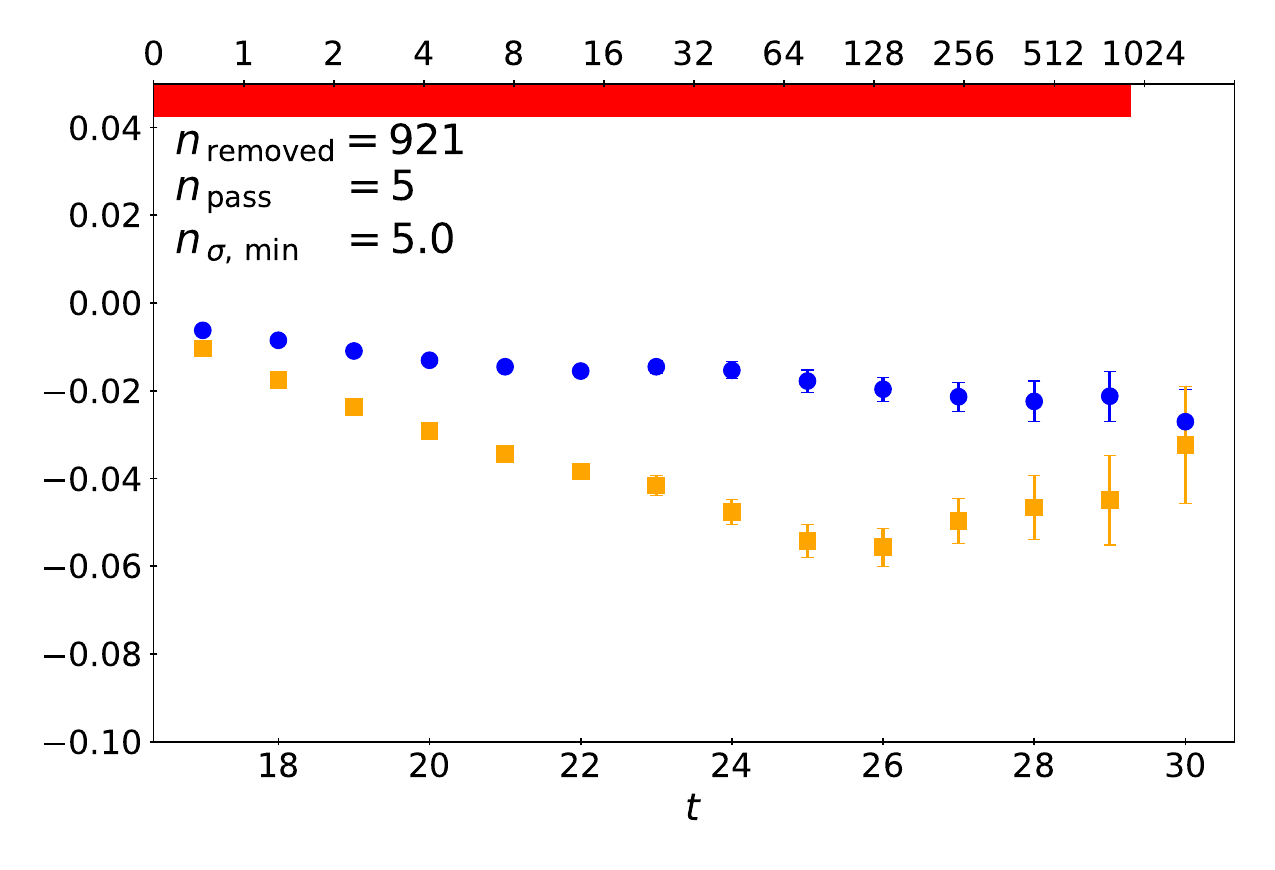}
    \includegraphics[width=0.3\linewidth]{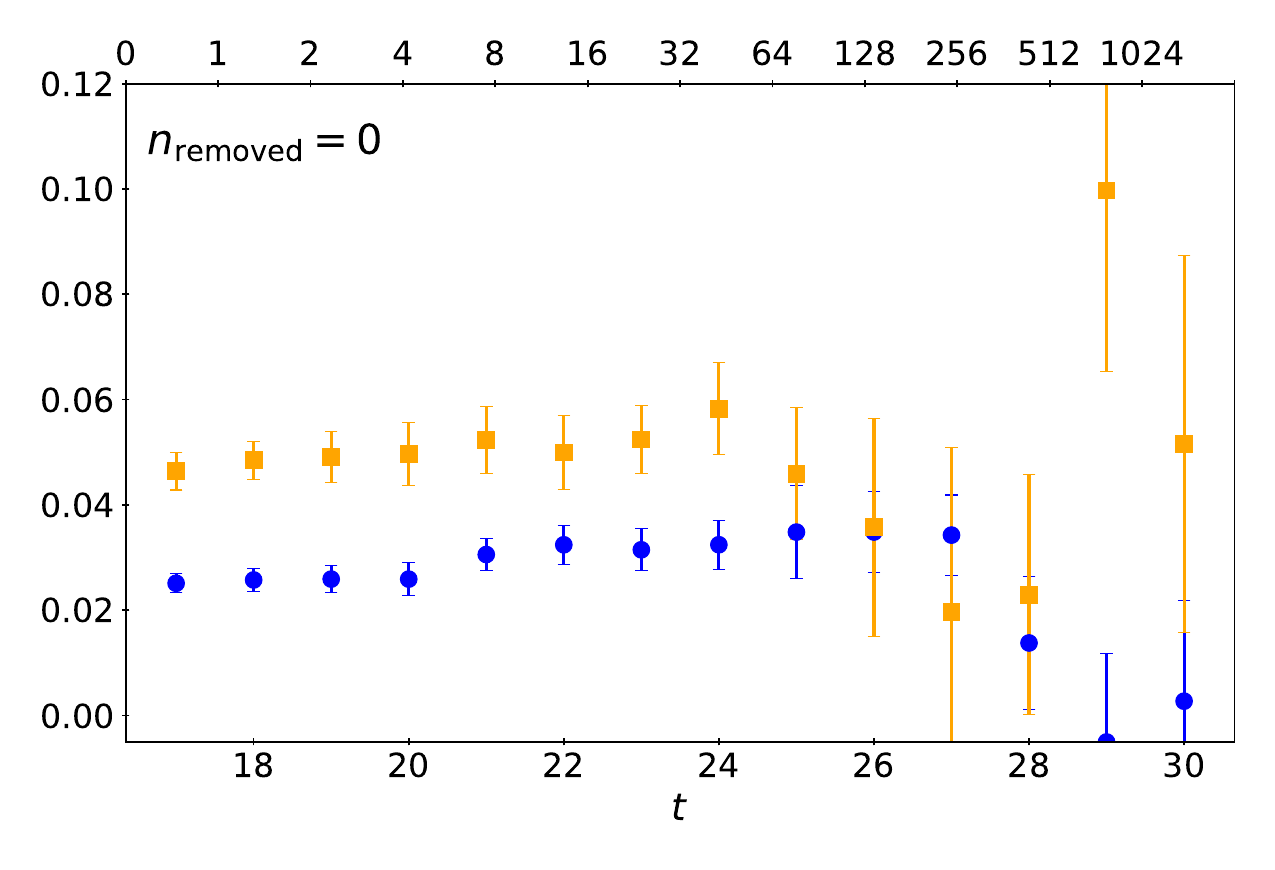}
    \includegraphics[width=0.3\linewidth]{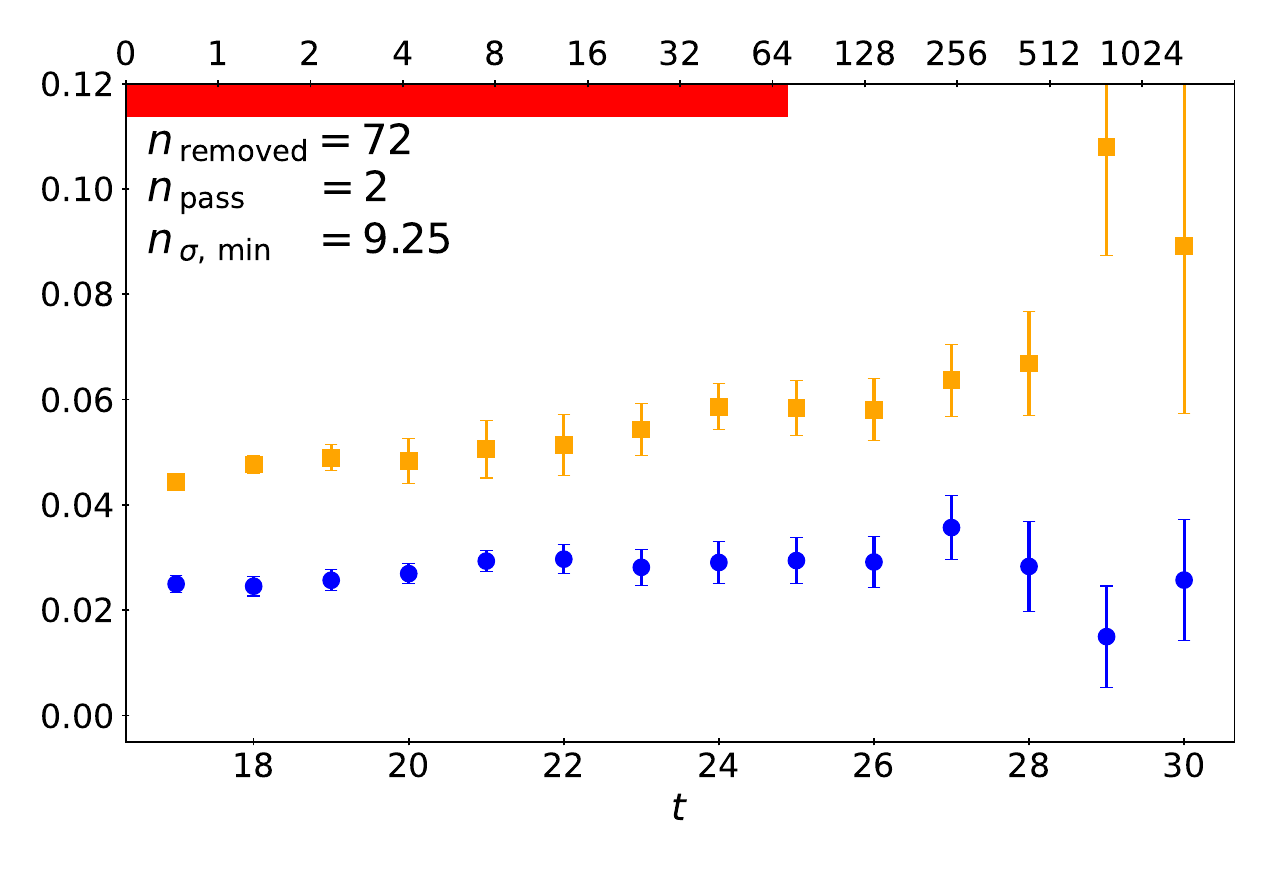}
    \caption{The effective energy shift $\delta E_{\beta}(B,t)$, of
        \autoref{eqn:polarisability:energy:ratio} for the \cascadez\ (top) and \cascadem\ (bottom).
        Plots and symbols are as in \autoref{fig:results:nucleonplateaus}. For the \cascadem\ in the
        bottom row, there is no noticeable jump in the $k_d=1$ plateau before the $k_d=2$ plateau is
        resolved. As such, we simply include the plot with no exceptional configurations removed on
        the left and the final $k_d=2$ plateau resolution on the right.}
    \label{fig:results:cascadeplateaus}
\end{figure*}

The plateaus for the $\Xi$ baryons are shown in \autoref{fig:results:cascadeplateaus}.
With the strange sector being doubly represented, the plateau profile of the $\Xi$ baryons is
cleaner again. The pattern of the $u$-quark causing increased susceptibility to exceptional
configurations continues for the \cascadez\ which requires removal of 921 configurations compared
to just 72 for the \cascadem. We do highlight that while the plateau for the \cascadez\ takes a
long time to fully improve, the plateau is extremely good once it does.

The plateau profiles are shown in \autoref{fig:results:cascadeplateaus} but it is important to note
here, that the decrease in susceptibility means that the clear jumps in quality of the effective
energy are less well defined. This is especially true of the $k_d=1$ plateau which improves much
more gradually, particularly for the \cascadem, the least susceptible baryon. For this reason, we
omit the intermediate plot for the \cascadem\ in \autoref{fig:results:cascadeplateaus} and simply
include the untouched effective energy and the final effective energy when the $k_d=2$ plateau is
resolved.

The quality of the untouched \cascadez\ plateau allows fitting in the range $[t_{\rm min},t_{\rm
    max}]=[20,25]$ as the rapidly increasing $k_d=2$ plateau prevents extending the fit later
without strongly biasing early, potentially contaminated windows. Due to the quality of the
\cascadem\ plateau, we are very aggressive, terminating the fit at the first sign of noise, hence
fitting in the range $[t_{\rm min},t_{\rm max}]=[20,24]$. With the removal of exceptional
configurations, we are able to significantly extend the fit range in both cases. We fit in the
range $[t_{\rm min},t_{\rm max}]=[20,29]$ for the \cascadez\ and $[t_{\rm min},t_{\rm
    max}]=[20,28]$ for the \cascadem.

% Windows for easy reference [untouched] [removed]
% cascade0: [20,25], [20,29]
% cascadem: [20,24], [20,28]

\section{Chiral extrapolation}\label{sec:chiralextrapolation}

\begin{figure*}
    \subfloat[\label{fig:chiral:protonextrap}]{
        \includegraphics[width=0.48\linewidth]{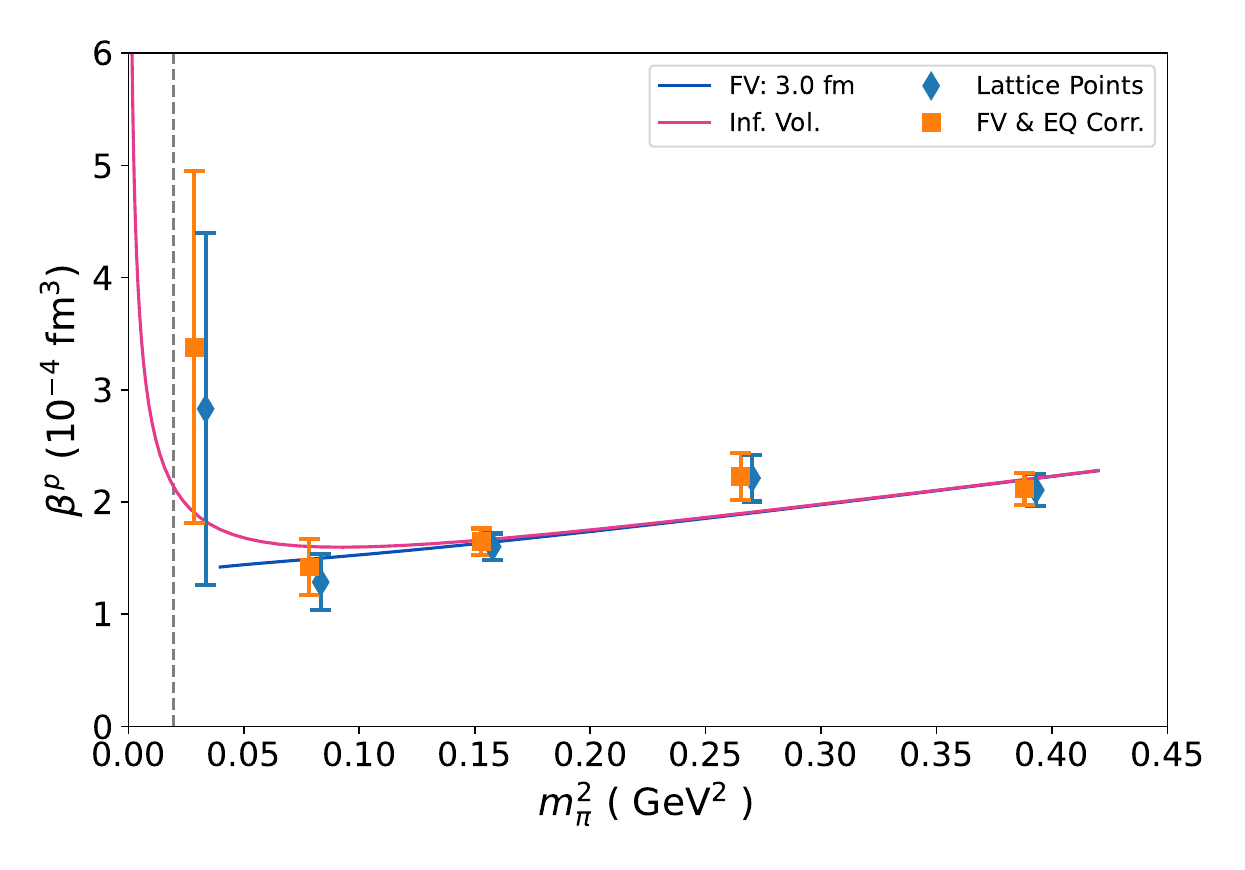}
    }
    \subfloat[\label{fig:chiral:neutronextrap}]{
        \includegraphics[width=0.48\linewidth]{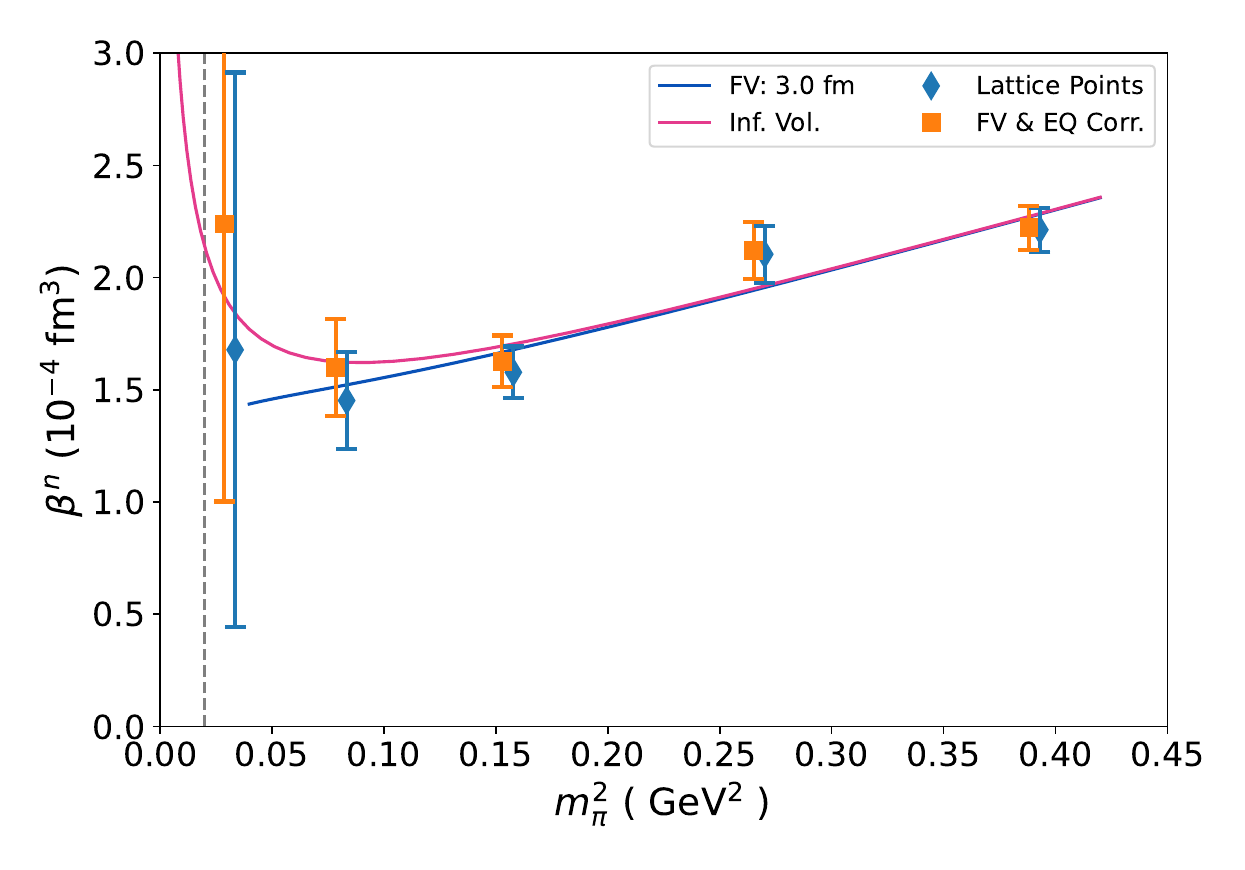}
    }\\
    \subfloat[\label{fig:chiral:simapextrap}]{
        \includegraphics[width=0.48\linewidth]{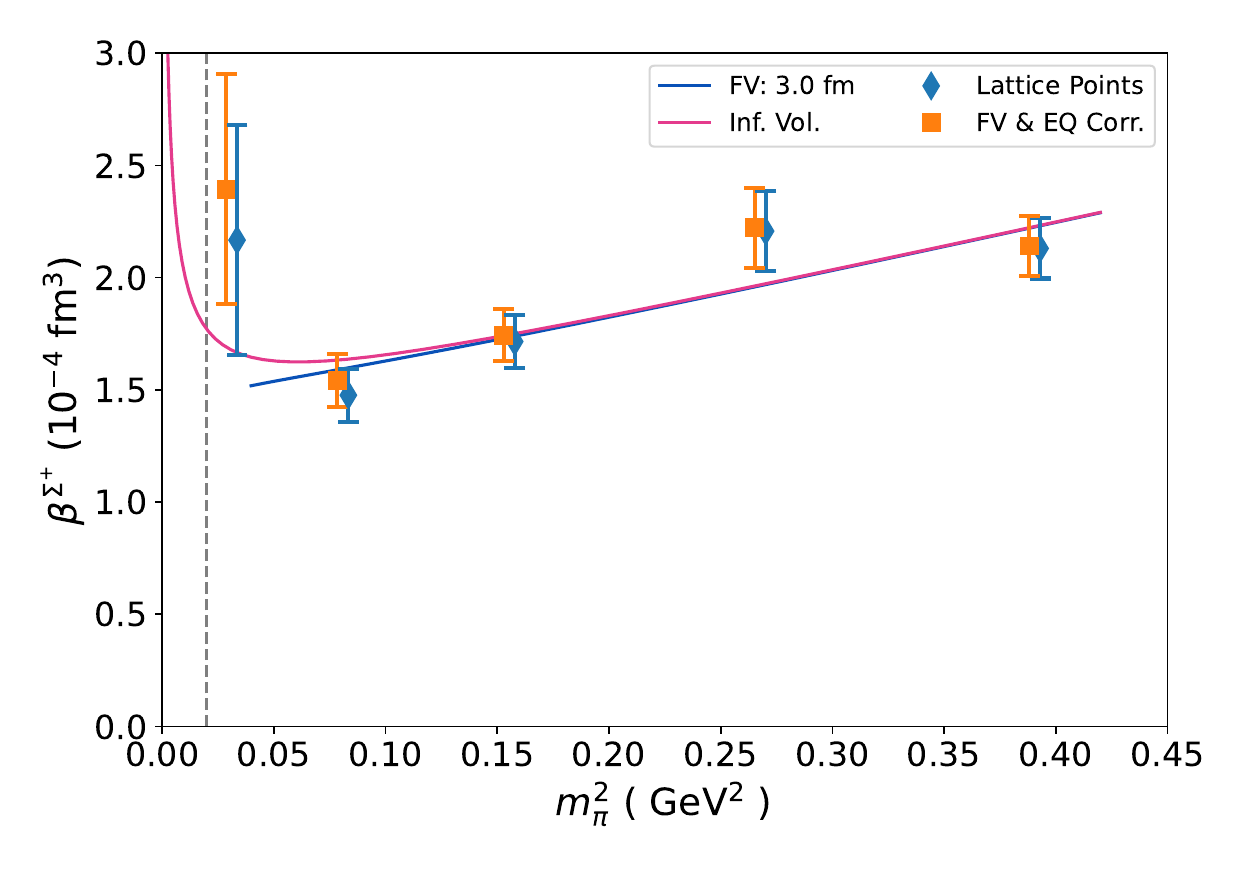}
    }
    \subfloat[\label{fig:chiral:sigmamextrap}]{
        \includegraphics[width=0.48\linewidth]{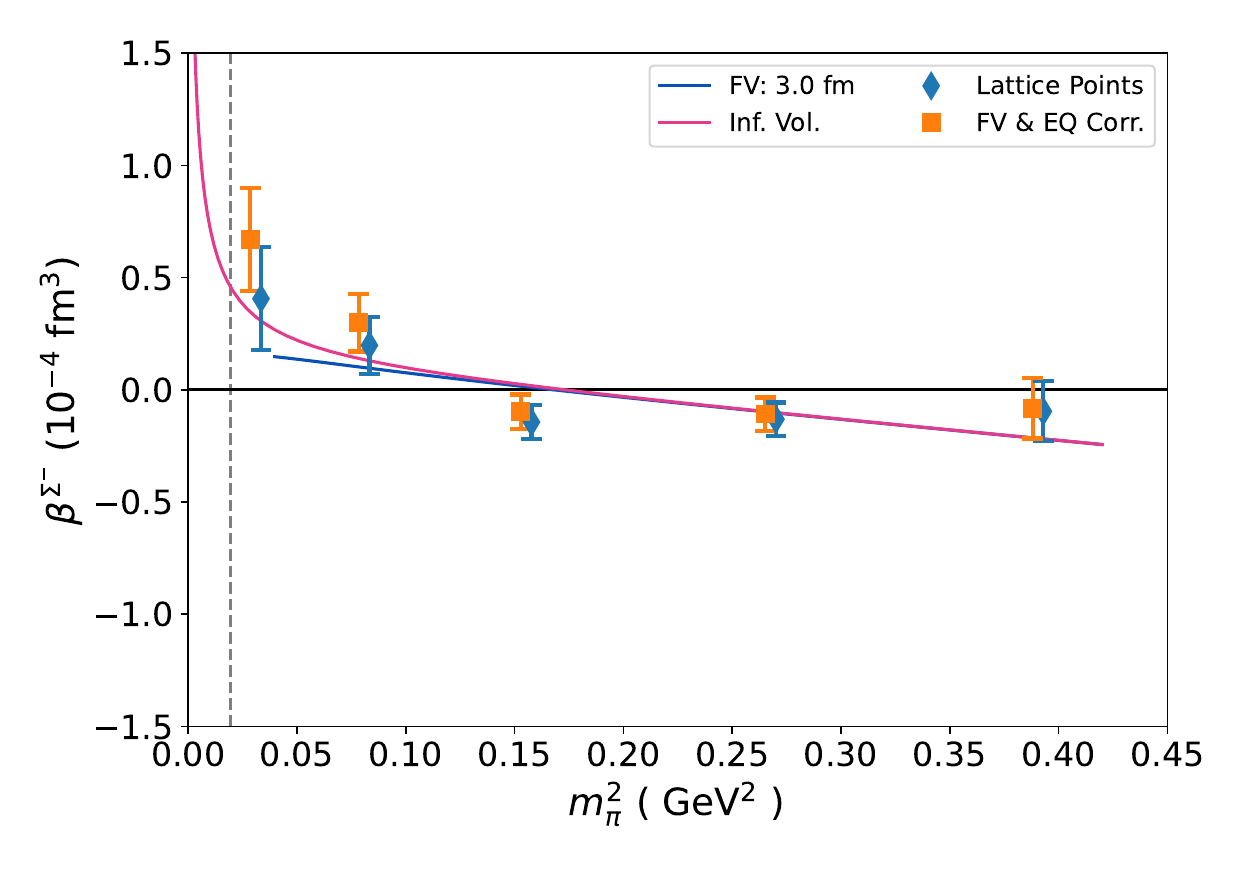}
    }\\
    \subfloat[\label{fig:chiral:cascade0extrap}]{
        \includegraphics[width=0.48\linewidth]{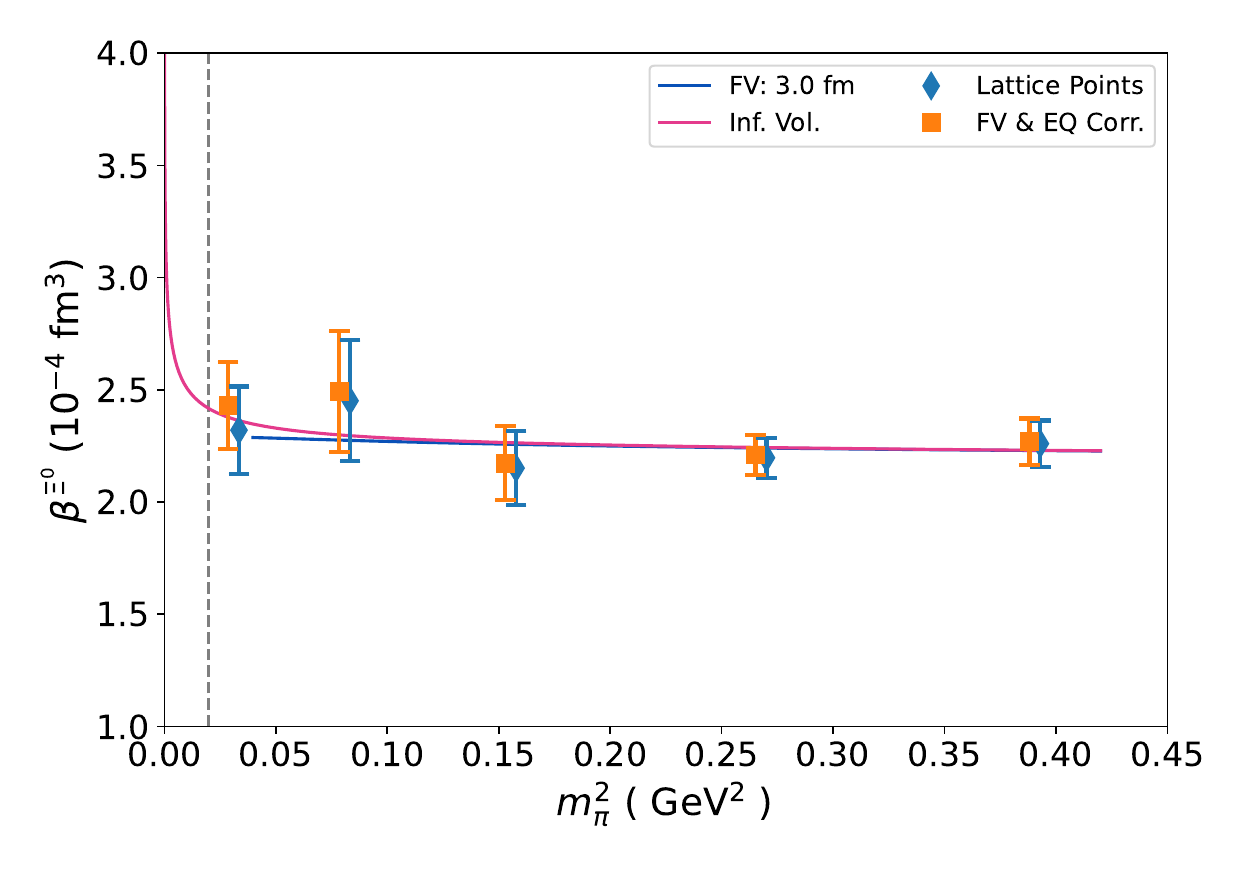}
    }
    \subfloat[\label{fig:chiral:cascademextrap}]{
        \includegraphics[width=0.48\linewidth]{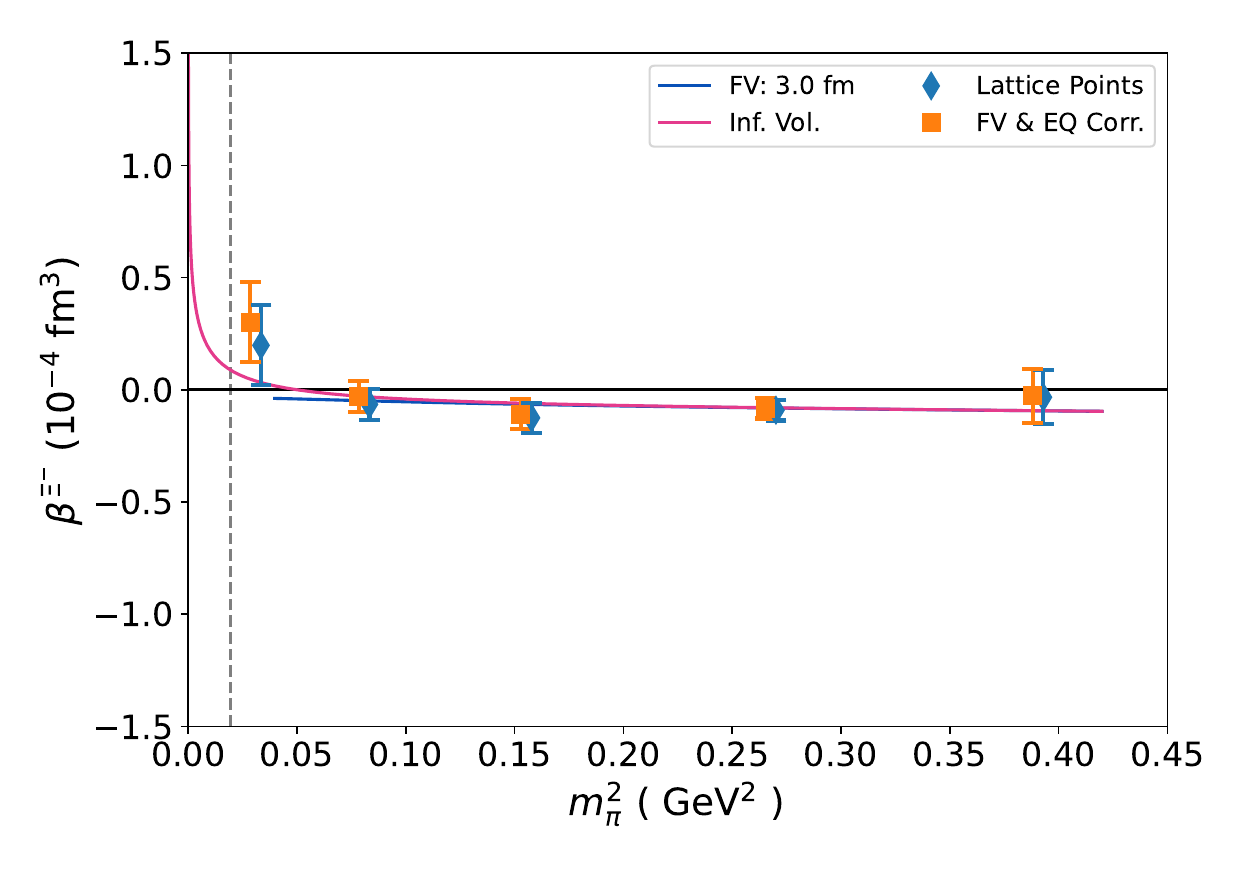}
    }
    \caption{Chiral extrapolation of the finite-volume and electro-quench-corrected (FV \ EQ
        Corr.) octet-baryon magnetic polarisabilities to the physical pion mass (vertical dashed
        line). Extrapolation fits for (a) $p$, (b) $n$, (c) \sigmap, (d) \sigmam, (e) \cascadez, and
        (f) \cascadem\ baryons are presented. Uncorrected lattice points (Lattice Points) are
        horizontally offset from the corrected points.  Extrapolation curves for the 3 fm finite
        volume for $m_\pi\, L > 3$ (FV. 3.0 fm), and the infinite-volume extrapolation (Inf.\ Vol.)
        are illustrated.  The methodology is described in Ref.~\cite{kabelitz:octet} where the values
        at the four heaviest masses originate. }
    \label{fig:chiralextrapolation:extrapolation}
\end{figure*}

In Ref.~\cite{kabelitz:octet}, we performed a chiral extrapolation to the physical pion mass which
includes a correction for the electro-quenching of the calculation and an estimate of the finite
volume correction. The corrections follow from a consideration of finite-volume partially-quenched
chiral effective field theory for baryon magnetic polarisabilities
\cite{Hall:2013dva,kabelitz:octet}.  The details of the corrections and subsequent extrapolation
are given there. Here we include the new light-quark-mass values calculated following the removal
of exceptional configurations.  Our aim is to provide context for the values determined in the
previous section.

The pion mass dependence observed for each baryon is illustrated in
\autoref{fig:chiralextrapolation:extrapolation}. In the figure, we include both the original
lattice QCD results and the values obtained after correcting the lattice QCD results for the
finite-volume and electro-quenching. The extrapolation is illustrated for both finite and infinite
volume.

We see that even after the removal of exceptional configurations, the uncertainty for the proton
magnetic polarisability at the lightest quark mass remains large. Still, its magnitude is correct
and the $1\sigma$ uncertainty is in accord with both finite-volume and infinite-volume
extrapolations, largely determined by the four heavier quark mass points.  Perhaps the most
important insight is an indication of the level of statistics required to resolve a value with an
uncertainty similar to the heavier quark mass considered.  Our four-fold increase in statistics
over the second-lightest quark mass is not yet sufficient.

Similar results are observed for the neutron.  The uncertainties are a little smaller, in accord
with a singly-represented $u$ quark, and the central values are in better accord with the
expectations from the extrapolation curves.  It is the presence of three light valence quarks that
presents a genuine challenge to precision determinations of the nucleon polarisabilities.

Significant improvement is observed for the hyperons.  Even a single strange quark within the
\sigmap\ is sufficient to tame the unwieldy uncertainty observed for the proton.  This time the
results sit a little high, just outside the $1\sigma$ uncertainty estimate. This trend to sit a
little high is also observed for the other charged hyperons. However, the agreement is at the
$1\sigma$ level.  Indeed, for the \sigmam, \cascadez, and \cascadem, we observe values obtained
following the removal of exceptional configurations that are in accord with the predictions of
chiral effective field theory.

The inclusion of values at the lightest quark mass in the chiral fit has minimal impact on the
result of the extrapolation to the physical point.
In light of the size of the uncertainties and the general agreement observed, this is as one might
expect.
As a result, we direct the reader to Ref.~\cite{kabelitz:octet} for our best physical predictions
of baryon magnetic polarisabilities, all obtained without the requirement of removing exceptional
configurations.

\section{Conclusion}\label{sec:conclusion}

Baryon magnetic polarisabilities have been calculated for the first time on the $m_\pi = 156$ MeV
PACS-CS ensemble \cite{PACS-CS2008ensembles}, closer to the physical point than in any previous
work. In pursuing the calculations, an exceptional configuration problem was
identified.

Considering the use of a Wilson-type fermion action with additive mass renormalisation,
the problem appears to be associated with the electro-quenching of the sea-quarks in the generation
of the dynamical-fermion configurations, without regard to the background fields introduced
herein. This problem is especially relevant for the nucleons which are most susceptible to such
configurations due to their light valence-quark composition. We observed that even a single strange
quark can have a strong effect in suppressing the influence of exceptional configurations on baryon
correlation functions.
Furthermore, the larger electric charge magnitude of the up quark is observed to further enhance
the susceptibility of a baryon correlation function to exceptional configuration problems.

Our main focus has been to devise an algorithm to identify and eliminate exceptional configurations
from baryon correlation functions at finite background magnetic field. A key difficulty encountered
is to develop a stopping criteria where one could be confident that the minimal number of
exceptional configurations are removed from the ensemble.  Here, animations of the correlation
function evolution as configurations are removed are particularly helpful, revealing a step change
in the resolution of the correlation functions, after which further evolution was slow.  Our criteria
is to stop exceptional configuration removal immediately after the observed step change.

Using this algorithm to systematically identify and remove exceptional configurations, baryon
magnetic polarisabilities are extracted for each of the outer octet baryons containing a
doubly-represented quark flavour.  Our omission of \sigmaz\ and $\Lambda$ is due to their mixing in
the presence of a background field.

The results are compared with the expectations of partially-quenched finite-volume choral effective
field theory.  Agreement is observed, typically at the $1\sigma$ level, with a trend for the new
light quark-mass results to sit a little high.  In light of the larger uncertainties that remain
even after exceptional configuration elimination and the general agreement observed with the
expectations of chiral effective field theory, the inclusion of these new lattice QCD results have
minimal impact on the predictions of the chiral extrapolation at the physical point.  To this end,
we defer to our previous publication \cite{kabelitz:octet} for our best physical predictions of
baryon magnetic polarisabilities, obtained without any requirement of removing exceptional
configurations.

To better determine the magnetic polarisability at light near-physical quark masses, there are
several avenues to explore.  The obvious approach is to generate new ensembles of dynamical-fermion
configurations including interactions between the electrically charged quarks and the background
magnetic field. This removes the electro-quenching issue, and should eliminate the presence of
exceptional configurations. However, such an approach loses the important QCD correlations withing
the correlation function ratios constructed to resolve the magnetic polarisability. The requirement
of an extraordinary increase in statistics is anticipated.

An interim approach, could exploit the improved chiral properties of a chiral fermion action, such
as the overlap fermion action.  In this case, renormalisation of the quark mass is multiplicative
and this should largely address the issue of exceptional configurations.  While the overlap action
is ${\mathcal O}(100)$ times computationally expensive, correlation functions would maintain the
aforementioned QCD correlations.  Given that partially-quenched chiral effective field theory
predicts the electro-quenching effects are a relatively small correction to the baryon magnetic
polarisabilities, this presents as a reasonable interim approach. \\

Baryon correlation functions were constructed using the COLA software library, developed at the University of Adelaide~\cite{COLA}. \\

The results presented herein are founded on the PACS-CS ensembles~\cite{PACS-CS2008ensembles}.  The
ensembles are created by the PACS-CS collaboration, Aoki {\it et al.} and are available
from~\cite{jldg}. The ensembles in question are part of the April 2009 release.

\begin{acknowledgments}
    We thank the PACS-CS Collaboration for making their $2+1$ flavour configurations available and the
    ongoing support of the International Lattice Data Grid (ILDG).
    WK was supported by the Pawsey Supercomputing Centre through the Pawsey Centre for Extreme Scale Readiness (PaCER) program. This work was supported with supercomputing resources provided by the Phoenix HPC service at the University of Adelaide. This research was undertaken with the assistance of resources from the National Computational Infrastructure (NCI), which is supported by the Australian Government. This research is supported by Australian Research Council through Grants No.~DP190102215 and No.~DP210103706.
\end{acknowledgments}

\bibliography{main}

\clearpage

\input{supplementary}

\end{document}

%% file: supplementary.tex
\title{Supplemental Material: Magnetic polarisability energy shift animations}
\begin{abstract}
    This supplementary document provides animations of the  energy shift $\delta E_{\beta}(k_d,t)$ of Eq.~\ref{eqn:polarisability:energyshift} for each of the outer octet baryons. As the animation evolves, exceptional configurations are removed. The red bar at the top of the plot indicates the number of exceptional configurations removed to achieve the resolution presented. This and the corresponding pass number and $n_{\sigma,\,{\rm min}}$ are also indicated. \\

    As in Ref.~\cite{Mickley:2024zyg}, the animations are produced using the \texttt{animate} package in \LaTeX. To play the animations, readers must open this document in any of Acrobat Reader, KDE Okular, PDF-XChange or Foxit Reader.

    The controls for the animations are located below their thumbnails. From left to right, these are: stop and go to first frame, step backwards one frame, play backwards, play forwards, step forwards one frame, and stop and go to last frame. Simply click on the desired control to interact with the animation. After clicking either play button, they will be replaced with a pause button which can subsequently be used to stop the animation.

\end{abstract}

\makeatletter
\@booleanfalse\preprint@sw
\makeatother
\maketitle

\setcounter{figure}{0}
\renewcommand\thefigure{S-\arabic{figure}}
\renewcommand{\theHfigure}{Supplement.\thefigure}

\begin{figure*}
    \centering
    \animategraphics[loop,nomouse,width=0.8\linewidth,controls={play,stop,step}]{5}{frames_proton_1}{}{}
    \caption{
        Animations of the  energy shift $\delta E_{\beta}(k_d,t)$ of Eq.~\ref{eqn:polarisability:energyshift} for the proton. As shown in \autoref{fig:results:nucleonplateaus} (top), the $k_d=1$ and $2$ plateaus are resolved with the removal of 283 and 773 exceptional configurations respectively.
    }\label{fig:protonanimation}
\end{figure*}

\begin{figure*}
    \centering
    \animategraphics[loop,nomouse,width=0.8\linewidth,controls={play,stop,step}]{5}{frames_neutron_1}{}{}
    \caption{
        Animations of the  energy shift $\delta E_{\beta}(k_d,t)$ of Eq.~\ref{eqn:polarisability:energyshift} for the neutron. As shown in \autoref{fig:results:nucleonplateaus} (bottom), the $k_d=1$ and $2$ plateaus are resolved with the removal of 30 and 60 exceptional configurations respectively.
    }\label{fig:neutronanimation}
\end{figure*}

\begin{figure*}
    \centering
    \animategraphics[loop,nomouse,width=0.8\linewidth,controls={play,stop,step}]{5}{frames_sigmap_1}{}{}
    \caption{
        Animations of the  energy shift $\delta E_{\beta}(k_d,t)$ of Eq.~\ref{eqn:polarisability:energyshift} for the \sigmap. As shown in \autoref{fig:results:sigmaplateaus} (top), the $k_d=1$ and $2$ plateaus are resolved with the removal of 254 and 680 exceptional configurations respectively.
    }\label{fig:sigmapanimation}
\end{figure*}

\begin{figure*}
    \centering
    \animategraphics[loop,nomouse,width=0.8\linewidth,controls={play,stop,step}]{5}{frames_sigmam_1}{}{}
    \caption{
        Animations of the  energy shift $\delta E_{\beta}(k_d,t)$ of Eq.~\ref{eqn:polarisability:energyshift} for the \sigmam. As shown in \autoref{fig:results:sigmaplateaus} (bottom), the $k_d=1$ and $2$ plateaus are resolved with the removal of 128 and 225 exceptional configurations respectively.
    }\label{fig:sigmamanimation}
\end{figure*}

\begin{figure*}
    \centering
    \animategraphics[loop,nomouse,width=0.8\linewidth,controls={play,stop,step}]{5}{frames_cascade0_1}{}{}
    \caption{
        Animations of the  energy shift $\delta E_{\beta}(k_d,t)$ of Eq.~\ref{eqn:polarisability:energyshift} for the \cascadez. As shown in \autoref{fig:results:cascadeplateaus} (top), the $k_d=1$ and $2$ plateaus are resolved with the removal of 129 and 921 exceptional configurations respectively.
    }\label{fig:cascadezanimation}
\end{figure*}

\begin{figure*}
    \centering
    \animategraphics[loop,nomouse,width=0.8\linewidth,controls={play,stop,step}]{5}{frames_cascadem_1}{}{}
    \caption{
        Animations of the  energy shift $\delta E_{\beta}(k_d,t)$ of Eq.~\ref{eqn:polarisability:energyshift} for the proton. As shown in \autoref{fig:results:nucleonplateaus} (top), the $k_d=1$ and $2$ plateaus are resolved with the removal of 72 exceptional configurations.
    }\label{fig:cascademanimation}
\end{figure*}